\documentclass{article}
\usepackage{PRIMEarxiv}

\usepackage[utf8]{inputenc} 
\usepackage[T1]{fontenc}    
\usepackage{hyperref}       
\usepackage{url}            
\usepackage{xcolor}         
\usepackage{amsfonts, amssymb, amsmath} 
\usepackage{graphicx}       
\usepackage{booktabs}       
\usepackage{multirow}       
\usepackage{subcaption}
\usepackage{adjustbox}
\usepackage{makecell}
\usepackage{tabularx}
\usepackage{algorithmic}
\usepackage{textcomp}
\usepackage{nicefrac}       
\usepackage{microtype}      
\usepackage{lipsum}
\usepackage{fancyhdr}       
\usepackage{threeparttable} 
\usepackage{biblatex}
\usepackage[english]{babel}
\usepackage{blindtext}
\usepackage{footmisc}
\usepackage[justification=centering]{caption} 

\def\BibTeX{{\rm B\kern-.05em{\sc i\kern-.025em b}\kern-.08em
    T\kern-.1667em\lower.7ex\hbox{E}\kern-.125emX}}

\newcolumntype{C}{>{\centering\arraybackslash}X} 
\definecolor{FinMem}{HTML}{d14749}
\definecolor{FinGPT}{HTML}{4e89e0}
\definecolor{Park}{HTML}{59a14f}
\definecolor{A2C}{HTML}{ee4199}
\definecolor{PPO}{HTML}{f28e2b}
\definecolor{DQN}{HTML}{8F337F}
\title{\textsc{FinMem}: A Performance-Enhanced LLM Trading Agent with Layered Memory and Character Design}

\author{ 
    \textbf{Yangyang Yu$^{*}$}, \textbf{Haohang Li$^{*}$}, \textbf{Zhi Chen$^{*}$}, \textbf{Yuechen Jiang$^{*}$}, \textbf{Yang Li$^{*}$} \\
    \textbf{Denghui Zhang}, \textbf{Rong Liu}, \textbf{Jordan W. Suchow}, \textbf{Khaldoun Khashanah\textdagger}\\
    \textit{Stevens Institute of Technology}\\ 
    Hoboken, NJ, United States \\
    \texttt{\{yyu44, hli113, zchen100, yjiang52, yli269, dzhang42, rliu20, jws\}@stevens.edu}\\
}
\date{}

\begin{document}
\maketitle


\def\thefootnote{\textdagger} \footnotetext{Corresponding author. Email: kkhashan@stevens.edu}\def\thefootnote{\arabic{footnote}}
\def\thefootnote{*}\footnotetext{Equal contribution, with author order decided by dice roll.}\def\thefootnote{\arabic{footnote}}
\def\thefootnote{2}\footnotetext{The source code of this project can be found via: \href{https://github.com/pipiku915/FinMem-LLM-StockTrading/} {\textsc{FinMem} LLM Trading}}\def\thefootnote{\arabic{footnote}}

\begin{abstract}
Recent advancements in Large Language Models (LLMs) have exhibited notable efficacy in question-answering (QA) tasks across diverse domains. Their prowess in integrating extensive web knowledge has fueled interest in developing LLM-based autonomous agents. While LLMs are efficient in decoding human instructions and deriving solutions by holistically processing historical inputs, transitioning to purpose-driven agents requires a supplementary rational architecture to process multi-source information, establish reasoning chains, and prioritize critical tasks. Addressing this, we introduce \textsc{FinMem}, a novel LLM-based agent framework devised for financial decision-making. It encompasses three core modules: Profiling, to customize the agent's characteristics; Memory, with layered message processing, to aid the agent in assimilating hierarchical financial data; and Decision-making, to convert insights gained from memories into investment decisions. Notably, \textsc{FinMem}'s memory module aligns closely with the cognitive structure of human traders, offering robust interpretability and real-time tuning. Its adjustable cognitive span allows for the retention of critical information beyond human perceptual limits, thereby enhancing trading outcomes. This framework enables the agent to self-evolve its professional knowledge, react agilely to new investment cues, and continuously refine trading decisions in the volatile financial environment. We first compare \textsc{FinMem} with various algorithmic agents on a scalable real-world financial dataset, underscoring its leading trading performance in stocks. We then fine-tuned the agent's perceptual span and character setting to achieve a significantly enhanced trading performance. Collectively, \textsc{FinMem} presents a cutting-edge LLM agent framework for automated trading, boosting cumulative investment returns. 
\end{abstract}

\textbf{KEYWODS}: Financial AI, Large Language Model, Trading Algorithms, Deep Learning, Financial Technology

\section{Introduction}

With the influx of diverse financial data streams from the web, traders face a deluge of information from various sources. This requires them rapidly to understand, memorize, and filtrate crucial events for investment decisions. However, innate cognitive limitations restrict human traders from processing information within their perception and memory capacity, a span much narrower than the actual volume of available information \cite{black1986noise}. Consequently, insufficiently considering or even dismissing critical events affecting trading decisions becomes increasingly concerning as data availability expands. To overcome the physical limitations in the memory systems of human traders, researchers have been consistently working on designing autonomous trading agent systems. These systems need to thoroughly integrate all available information and possess a sophisticated design in the agent's backbone algorithm to deliver enhanced trading performance.

The evolution of autonomous trading systems has transitioned from the initial rule-based trading strategies \cite{edwards2018technical} to more advanced machine-learning-based algorithms \cite{huang2019automated}. In recent years, Reinforcement Learning (RL)-based agents \cite{fischer2018reinforcement}, especially those employing Deep Reinforcement Learning (DRL) \cite{millea2021deep} as backbone algorithms, garner joint attention of both academia and industry. Leveraging both RL principles and deep learning, DRL agents effectively handle and learn from scalable and diverse financial data, including stock prices, key financial indicators, and market sentiments. They utilize deep neural networks to extract expressive features from input data, representing the complex financial market environment, which enhances their comprehension ability. Retaining the key features of RL agents, they learn through interaction with a predefined environment to maximize investment gain over time. Research suggests DRLs can meet the crucial needs of trading agents to process and make informed decisions from large volumes of data. However, certain inherent features of DRL algorithms exhibit notable deficiencies in financial applications. \textbf{Firstly, DRL agents exhibit a lack of interpretability concerning the rationale behind their decisions \cite{balhara2022survey}.} They are often described as “black boxes,” where the internal processes and computational layers leading to a specific decision are neither easily understandable nor transparent. \textbf{Secondly, DRL agents find it challenging to effectively integrate textual data with numerical features.} Text data plays a vital role in finance, since the majority of market information is conveyed through news articles and financial reports. However, transforming text data to embeddings considerably increases input space dimensionality, making learning more computationally demanding. Plus, empirical studies have shown that combining textual representations with numerical financial indicators often leads to convergence challenges \cite{gershman2020neurobiology}. Therefore, a backbone algorithm with transparent reasoning and the enhanced ability to capture investment-related textual insights comprehensively is essential.


Recent advancements in Large Language Models (LLMs), like Generative Pre-trained Transformers (GPTs) \cite{openai2023gpt4}, offer viable solutions for developing trading agents, alleviating previous concerns. With carefully designed prompts, the LLM-based agent is able to provide reasons and outcomes in plain text. This allows for immediate observation and prompt adjustment of its reasoning process. Employing LLMs as the backbone algorithms for agents also overcomes the constraint of isolated environments. This is achieved through their vast pre-existing knowledge and the effective integration of valuable insights from a variety of data sources, including both textual and numerical. When equipped with suitable prompt templates, this approach significantly enhances decision-making capabilities \cite{wang2023survey}. Studies indicate that prompt-guided reasoning significantly improves problem-solving rates across various domains \cite{li2023making}. Notably, a growing body of research has focused on utilizing LLMs to make informed trading decisions for stocks and funds by continuously interacting with financial environment information \cite{yang2023fingpt, wu2023bloomberggpt}. However, in currently available approaches, LLMs primarily serve as a QA role rather than functioning as autonomous agents. \textbf{The potential issue with these approaches is the incapability of fully understanding the varying timeliness associated with different types of financial data.} These financial LLM agents, despite outperforming traditional trading benchmarks, generally process information indiscriminately through QA iterations, lacking the ability to memorize influential messages. Furthermore, their method of acknowledging the timeliness of financial data is heavily dependent on the uncertain and laborious LLM fine-tuning process. These insufficiencies undermine their ability to update the knowledge base in a daily manner, meaning they lack a memory component. As a result, they may struggle to prioritize significant and influential memory events effectively. Additionally, current literature on LLM-based trading agents lacks a comparative analysis between these applications and other autonomous trading systems, such as DRL agents.

To bridge this gap, we present \textsc{FinMem}, an innovative LLM-based autonomous trading agent with a novel layered memory system and dynamic character design. Unlike previous LLM agents in finance, \textsc{FinMem} encompasses a memory module adept at processing multi-source financial data with varying timeliness and self-adaptive character setting for fitting into volatile market environments.

Our concept is initially inspired by the Generative Agents framework by Park et al. \cite{10.1145/3586183.3606763}, aimed at enhancing the efficient retrieval of key events for general-purpose LLM agents. This framework features a unique character design and seed memory, activating the agent upon specific query through prompts. It prioritizes events in a unified memory stream, ranked by a linear combination of recency, relevancy, and importance. The framework outlined in \cite{10.1145/3586183.3606763} provides a foundational structure for LLM agent design. It includes a profiling module for character definition, a memory module for experience recording and critical information retrieval, and an action module to guide actions based on the retrieved memories. This structure effectively facilitates goal achievement for the agent in a general social environment. \textbf{However, Park et al.'s framework struggles with comprehending financial data with varying timeliness and importance}, like daily news versus quarterly and annual reports. Key challenges involve quantifying the timeliness of different information sources, optimizing information retrieval, and enhancing trading decisions with detailed analysis. To tackle these challenges, we further propose \textsc{FinMem} with the following improvements.

\textsc{FinMem} maintains a modular approach similar to Park et al. \cite{10.1145/3586183.3606763}, but features novel design of profiling and memory modules. Its specialized profiling module equips \textsc{FinMem} with a trading-task-specific professional background, enhancing robustness to market fluctuations via offering self-adaptive risk inclination option. \textsc{FinMem}'s memory module innovatively incorporates working memory and layered long-term memory components, ideal for stratified information processing.
Its working memory acts as a dynamic “workspace,” enabling operations like summarization, observation, and reflection on multi-source information to facilitate trading decisions. Its long-term memory, structured into shallow, intermediate, and deep layers \cite{craik1972levels}, manages varied decay rates to satisfy the need to retain distinct types of financial information within different time scales based on their corresponding timeliness. For instance, daily news, with its immediate effects on stock markets, is channeled into the shallow processing layer. Meanwhile, annual company reports, exerting a more prolonged impact, are processed in the deep layer by \textsc{FinMem}. Each layer in \textsc{FinMem} prioritizes memory events based on the assemble of recency, relevancy, and importance close to Park et al.'s method. However, it introduces new measurements for recency and importance, specifically tailored to better rank financial data according to their unique time sensitivity. \textsc{FinMem}'s memory mechanism can also transit significantly impactful investment memory events to deeper processing layers, ensuring their retention for extended periods. \textsc{FinMem}'s memory module can mirror the human cognitive system \cite{sweller2012human} and facilitate agile, real-time decisions \cite{sun2004desiderata}. It enables continuous evolution in professional knowledge through structured summarizing, retrospecting past experiences, and reacting to new trading scenarios. Additionally, \textsc{FinMem} includes a decision-making module capable of deriving investment decisions by considering top-ranked memory events and current market conditions.

Through experiments, we show that \textsc{FinMem} exhibits an outstanding ability to stratify and leverage the various levels of market insights, significantly improving the quality of trading decisions. We claim that \textsc{FinMem} provides these key contributions:

\textbf{\textsc{FinMem} presents a state-of-the-art LLM-based trading agent with a human-aligned memory mechanism and character design}, particularly crafted to capture investment insights from the financial market. In its agent memory module design, \textsc{FinMem} innovatively emulates human working and layered long-term memory mechanisms. This approach effectively harnesses the time-sensitive aspects of financial data, capturing crucial investment insights and thereby boosting trading performance. \textsc{FinMem}'s profiling module includes a dynamic character setting feature, offering seed information about professional backgrounds and adjustable risk inclinations. Additionally, it continuously updates domain knowledge as the trading experience grows, thereby enhancing \textsc{FinMem}'s capabilities. Ablation studies demonstrate that \textsc{FinMem} can learn from past trading experiences and evolve its knowledge base through continuous market interaction, maintaining robustness in complex markets for profitable trading decisions.

\textbf{\textsc{FinMem} can utilize its distinctive features to expand the agent's perceptual range beyond the human limitation to make well-informed trading decisions.} Cognitive research indicates that human working memory can recall only five to nine events at a time \cite{miller1956magical}. This limitation, while preventing information overload, can yield insufficient insight for precise decision-making. In contrast, \textsc{FinMem}'s memory module transcends this constraint. It allows adjusting cognitive load by selecting a flexible number of top-ranked events from each layer of its hierarchical long-term memory, allowing \textsc{FinMem} to maintain agility and deliver superior trading decisions in data-rich contexts.

\textbf{\textsc{FinMem} achieves impressive trading performance using training data that is limited in volume and spans a short time period.} Our experiments indicate that training \textsc{FinMem} with daily collected data over just six months to a year is enough to produce robust and notable trading results. This timeframe is considerably shorter than what other comparable models require. This efficiency is achieved through the optimal utilization of multi-source incoming data and the precise identification of key signals for trading decisions. Furthermore, it's noteworthy that \textsc{FinMem}'s effectiveness is demonstrated on smaller datasets and with general-purpose LLMs. Its capabilities are anticipated to be further amplified with access to larger, higher-quality financial datasets and LLMs fine-tuned specifically for financial applications.

In this paper, we begin by explaining the three core modules of \textsc{FinMem}. Subsequently, we emphasize its superior trading performance compared to a range of representative algorithmic agents. We further explore how \textsc{FinMem} achieves its optimal performance, examining adjustments in three key aspects: backbone algorithms, working memory capacity, and character settings.

\section{Related Work}
\subsection{Backbone Algorithms of Contemporary Autonomous Trading Agents}
The development of trading agents has evolved over several decades, influenced by advancements in technology, finance, and computational methodologies. Conventionally, a rule-based algorithm for trading stocks is an automated strategy that operates based on a predefined set of rules\cite{chen2012forecasting, vaidya2020moving, patari2014performance}. These rules are often derived from historical market patterns and trading experience. Compared with rule-based algorithms that use predefined rules and conditions, Reinforcement Learning provides a way for agents to learn by interacting with an environment and receiving feedback in the form of rewards or penalties. \cite{dang2019reinforcement, jangmin2006adaptive}. Deep learning models can be integrated with RL to handle large and complex state spaces, like those in stock markets. Such models are often referred to as Deep Reinforcement Learning (DRL) \cite{wu2020adaptive, xiong2018practical}. For example, Deep Q-Network (DQN) \cite{shi2021stock}, Advantage Actor-Critic (A2C) \cite{yang2020deep}, and Proximal Policy Optimization (PPO) \cite{liu2020finrl} are popular algorithms for such tasks. Using DRL agents as automated financial trading backbones, face two key issues: 1) A lack of interpretability, as their decisions, rooted in complex computations and high-dimensional representations, are challenging to articulate \cite{balhara2022survey}. 2) They struggle to fully leverage textual financial information due to the high-dimensional nature and computational intensity of rich text embeddings \cite{devlin2018bert, ethayarajh2019contextual}. Consequently, DRL agents often rely on extracting textual sentiment \cite{pricope2021deep}, sidestepping the direct use of embeddings \cite{chen2021sentiment, avramelou2023deep}, leading to an incomplete representation of crucial market information embedded in news and macroeconomic policies.

\subsection{Advancements from LLMs to LLM Autonomous Agents}
The evolution of LLMs has reshaped artificial intelligence and natural language processing. From foundational embeddings like Word2Vec \cite{goldberg2014word2vec} and GloVe \cite{pennington2014glove}, the field advanced with the introduction of sequential modeling like Long Short-Term Memory (LSTM) \cite{hochreiter1997long} and early transformer models of Bidirectional Encoder Representations from Transformers (BERT) \cite{devlin2018bert}. Today, the new-generation LLMs, like Generative Pre-trained Transformer series (GPTs) \cite{radford2018improving, openai2023gpt4} and LLM Meta AI (Llamas) \cite{touvron2023llama}, stand out in diverse QA tasks. The trend leans towards LLM agents.  While LLM agents for domain-specific tasks have been extensively researched \cite{huang2023memory, liffiton2023codehelp, park2022social}, their application in financial trading remains underexplored. Existing studies in this domain, such as \cite{wu2023bloomberggpt, yang2023fingpt}, often lack open-source availability or have not considered an architecture specifically tailored to fit the unique environment of finance markets. Thus, there's significant value in further investigating advanced, transparent LLM agents for trading.

\subsection{Architecture Design of LLM Autonomous Agent}

As Wang et al.\cite{wang2023survey} emphasizes, an effective architecture for LLMs serving autonomous agents is essential. Typically, this structure comprises modules like profiling, memory, planning, and actions, though not all may be essential for every application. There are cases of two modules (e.g., planning and action modules)  being integrated as one component (\cite{10.1145/3586183.3606763}). The design variations are numerous. For instance, profiling has been achieved through methods like handcrafting \cite{zhang2023building}, generation by LLMs \cite{wang2023recagent}, and alignment with real-world datasets \cite{argyle2023out}. Among these modules, the memory component is essential. Acting as the operational core, it aligns an agent's actions with real-world tasks. Research indicates that leveraging insights from cognitive science studies on human memory \cite{wang2023survey, sumers2023cognitive} can enhance this alignment. Thus, a well-structured trading LLM agent, comprising aptly designed modules, can sharply tackle the complexities of financial markets to make informed decisions.

\section{Architecture of \textsc{FinMem}}
\label{architecture}

In this section, we comprehensively detail the three core components of \textsc{FinMem}, namely the profiling, memory, and decision-making modules. The profiling module empowers \textsc{FinMem} to adaptively tailor character setting for specific trading tasks. Memory module leverages diverse time-efficiency attributes of financial data, enhancing its trading efficacy. The decision-making module enables \textsc{FinMem} to synchronize its memory streams with market facts, facilitating high-quality trading decisions. The details and notations associated with these three modules are provided in the subsequent sections.

\begin{figure*}[htbp]
\centering
\includegraphics[width=\columnwidth]{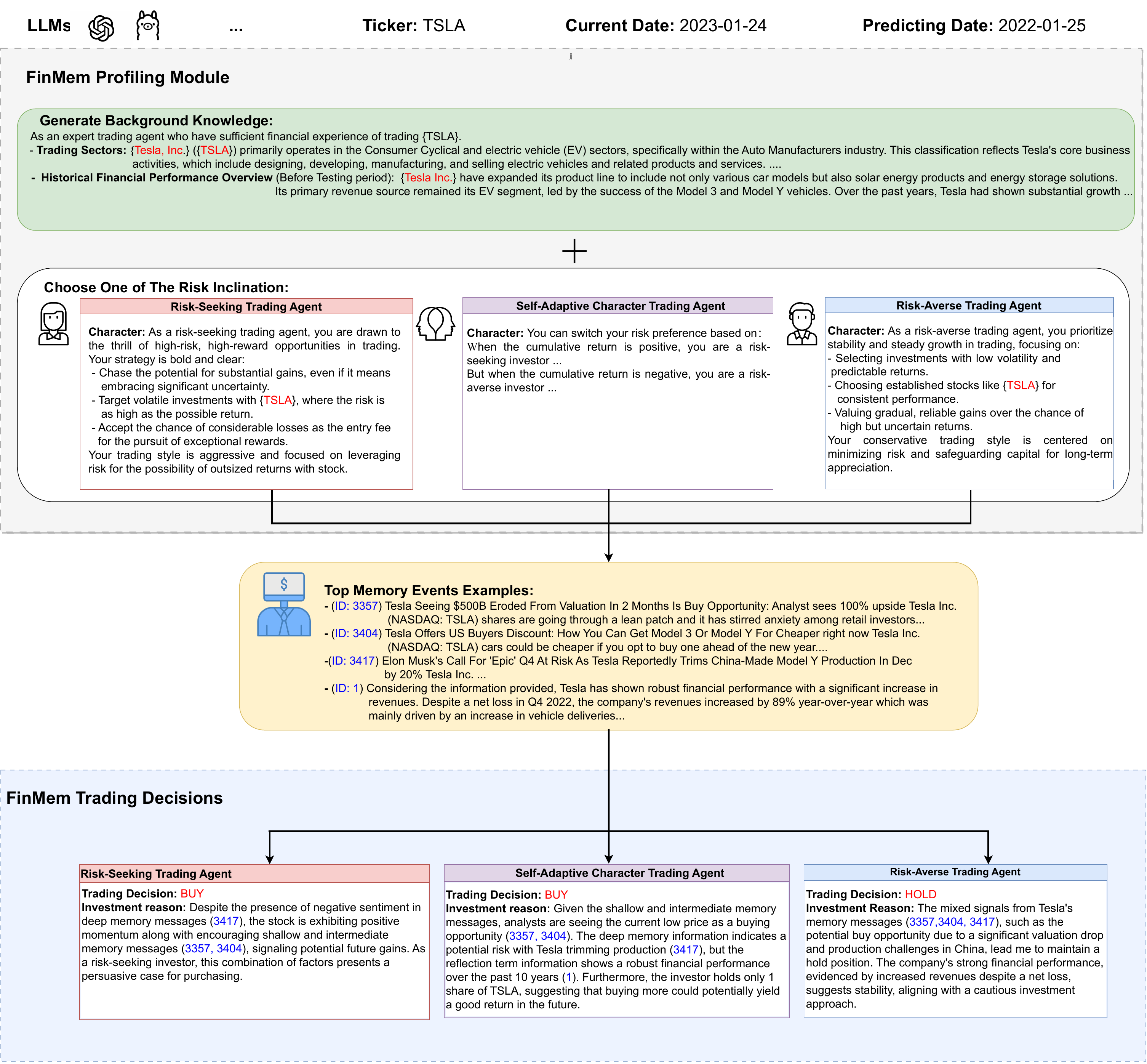}
\caption{The prompt template for \textsc{FinMem}'s profiling module. It includes two key elements of its character setting: professional background knowledge and three distinct investment risk inclinations. In the self-adaptive risk inclination option, the omitted texts align with the detailed descriptions provided for the risk-seeking and risk-averse inclinations.} 
\label{fig3:character_design}
\end{figure*}

\subsection{Profiling Module}
\label{profile_form}
The profiling module empowers \textsc{FinMem} to develop a dynamic agent character specifically designed to navigate the complex dynamics of financial markets effectively.

The dynamic character of \textsc{FinMem} comprises two principal components, as depicted in Figure~\ref{fig3:character_design}: firstly, a foundational professional knowledge base akin to a trading expert, and secondly, an agent with three distinct investment risk inclinations. The first component includes two types of information: an introduction to the primary trading sectors relevant to the company stock \textsc{FinMem} will trade in, and a concise overview of the historical financial performance of the specified ticker, spanning from the beginning to the end of the training period. Before initiating trades in a new company's stock, \textsc{FinMem} accesses and updates this sector-specific and historical financial data from a backend database. This professional background setting narrows down information and memory events pertinent to specific trading tasks.

The second component of \textsc{FinMem}'s design, illustrated in Figure~\ref{fig3:character_design}, encompasses three distinct risk inclination options: risk-seeking, risk-averse, and a self-adaptive risk character. The risk-seeking setting gears \textsc{FinMem} towards an aggressive, high-reward approach, while the risk-averse setting gears it towards a conservative, lower-risk strategy. A distinctive aspect of \textsc{FinMem} is its ability to dynamically alternate between these risk settings in response to current market conditions. Specifically, it shifts risk preferences when the Cumulative Return falls to below zero within a brief period, such as three days, and reversely. This flexible design functions as a protective mechanism, mitigating prolonged downturns in turbulent market environments. During the initial stage of the training phase, \textsc{FinMem} is configured with a chosen risk preference, each supplemented with comprehensive textual explanations through LLM prompts. These guidelines shape how \textsc{FinMem} processes incoming messages and determines its subsequent actions in alignment with its designated risk inclination. The system maintains a catalog of all risk inclinations and their detailed explanations in a backlog, enabling seamless adaptation to different stocks by switching among these risk profiles as needed.

The dynamic character setting in \textsc{FinMe}'s profiling module provides subjective and professional background knowledge and flexible choice of risk inclinations. It provides crucial context for filtering and retrieving trading-relevant information and memory events, thus improving accurate inferencing and adaptability to fluctuating market conditions.

\subsection{Memory Module}
\label{mem_form}
\begin{figure*}[htbp]
\centering
\includegraphics[width=\columnwidth]{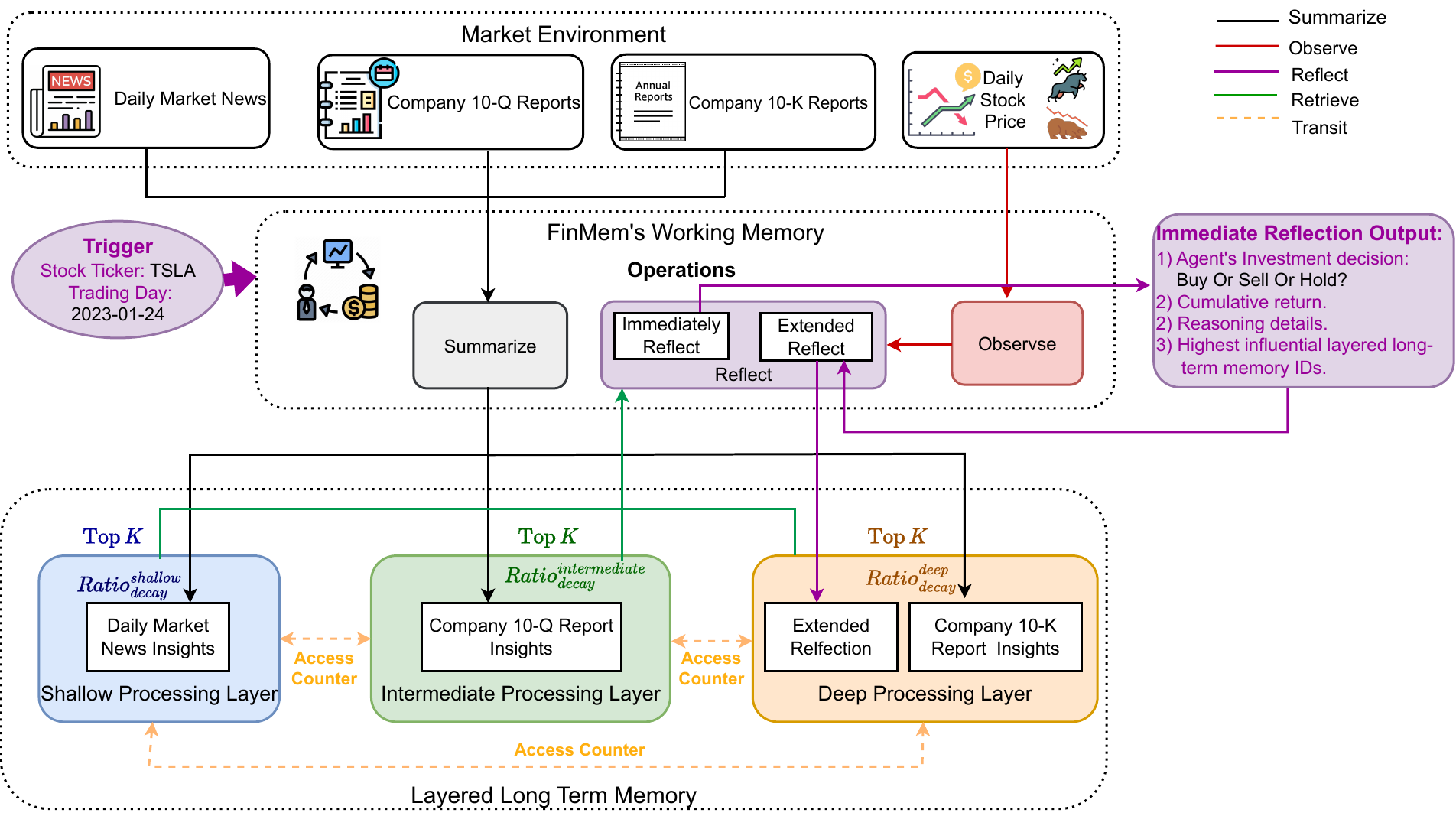}
\caption{Memory module structure of \textsc{FinMem} with a detailed view of components, operations, and workflow. The cognitive architectures of \textsc{FinMem}'s memory module  have two core components -- Working Memory and Layered Long-term Memory.} 
\label{fig1:training_test_logic}
\end{figure*}

The memory module of \textsc{FinMem} emulates a human trader's cognitive system so that it can efficiently process hierarchical financial information and prioritize the critical messages for high-quality investment decisions. Furthermore, it adjusts the memory span flexibly, enabling the agent to operate on a wider range of events over a longer retrieval period. \textsc{FinMem}'s memory module, illustrated in Figure~\ref{fig1:training_test_logic}, comprises working and long-term memory with layered processing capability and is initiated by a specific investment inquiry.

\subsubsection{Working memory}
\label{working memory}
\begin{figure*}[htbp]
\centering
\includegraphics[width=\columnwidth]{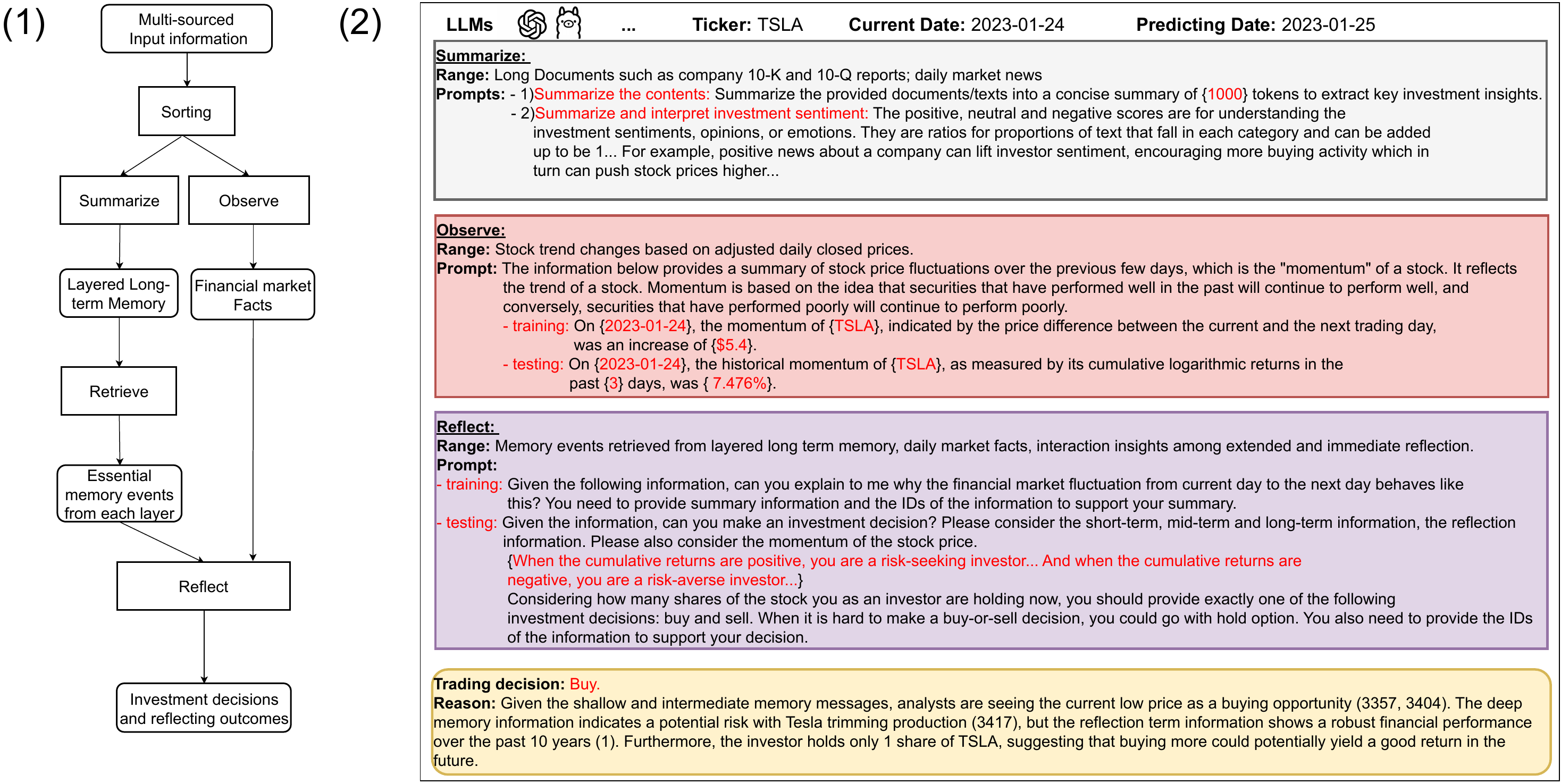}
\caption{(1) The decision-making module workflow of the \textsc{FinMem} trading agent retrieves critical memory events to inform specific decisions. (2) LLM prompt template used by \textsc{FinMem} to interact with incoming financial information.} 
\label{fig2:workflow}
\end{figure*}

Working memory refers to the human cognitive system's functions for temporary storage and diverse operations. We incorporate this concept into \textsc{FinMem}'s memory module development, creating a central workspace for informed decision-making. Unlike human working memory, having a maximum capacity of seven plus or minus two memory events \cite{miller1956magical}, \textsc{FinMem} has the ability to expand the capacity based on specific requirements. Tailored for converting financial data into trading actions, \textsc{FinMem}'s working memory encompasses three key operations: summarization, observation, and reflection. The mechanisms by which they interact and operate as an integrated decision-making workflow are detailed in the middle box of Figure~\ref{fig1:training_test_logic}. Additionally, the LLM prompt template that underpins these processes is thoroughly outlined in \ref{fig2:workflow}.

\textbf{Summarization:} \textsc{FinMem} leverages external market data to derive critical investment insights and sentiments tailored to specific stock trading queries, such as “Can you make an investment decision on TSLA on 1/24/2023?”. As illustrated in Figure~\ref{fig2:workflow} (2), this system condenses the original text into a compact yet informative paragraph, thereby enhancing \textsc{FinMem}'s processing efficiency. It efficiently extracts and summarizes pertinent data and sentiments for stock investment decisions, demonstrated here using Tesla Inc. as an example. Subsequently, \textsc{FinMem} directs these insights to an appropriate layer within its long-term memory architecture, selecting the layer based on the time sensitivity of the information.

\textbf{Observation:} Triggered the same inquiry, \textsc{FinMem} initiates an observation operation to gather market facts. The information available to \textsc{FinMem} varies between the training and testing phases. 

During the training phase, \textsc{FinMem} has access to comprehensive stock price data within the specified period. Upon receiving trading inquiries that specify a stock ticker and date, \textsc{FinMem} focuses on the daily adjusted closing price differences, comparing the following day's price with the current day's. These price differences are utilized as market ground labels. Specifically, a decrease in price suggests a “Sell” action, while an increase or no change in price indicates a “Buy” action.

During the testing phase, at a specific time point, \textsc{FinMem} loses the ability to access future price data. Its focus shifts to the analysis of historical stock price movements, depending on a retrospective evaluation of the cumulative return from the last $M$ trading days to infer future market trends. This phase, characterized by the absence of foreseen market grounds, serves as a critical assessment of \textsc{FinMem}'s development. It tests whether the system has adequately established logical connections between stock price trends and various financial information sources, such as news, reports, and indicators. This stage is key in evaluating \textsc{FinMem}'s capability of independently evolving its trading strategies for subsequent tasks, leveraging its analysis and interpretation of historical data patterns.

\textbf{Reflection:} Two types of reflections exist, immediate and extended reflection. (a) Immediate reflection is activated upon receiving a daily trading inquiry for a specific ticker. Using LLM and specific prompts exemplified in Figure~\ref{fig2:workflow} (2), the agent merges market indications and top-$K$-ranked events from each long-term memory layer. Market indications are derived from the outcomes of the observation operation and differ between the training and testing phases. During testing, this process yields three types of outputs: the trading direction (“Buy”, “Sell”, or “Hold”), the underlying rationale for this decision, and the most influential memory events, along with their IDs from each layer that informed the decision. In the training phase, specifying the trading direction is unnecessary, as \textsc{FinMem} is already informed of future stock movement directions. The top-$K$-ranked memory events encapsulate key insights and sentiments derived from critical investment-related incoming messages, all distilled by \textsc{FinMem}'s advanced summarization capabilities. 

(b) Extended reflection reevaluates immediate reflection outcomes for a ticker over a specified $M$-day trace period. It encompasses data like stock price trends, trading returns, and action rationales from multiple immediate reflections. While immediate reflection enables direct trading execution and records current feedback, extended reflection summarizes market trends and reassesses recent Cumulative Return on investment. Extended reflection is eventually transmitted and stored in the deep processing layer to emphasize its criticality (detailed introduced in Section~\ref{layered_long}) of long-term memory. $K$ and $M$ are hyperparameters to adjust \textsc{FinMem}'s working memory capacity and information retrieval ability. \textsc{FinMem} gains the flexibility of integrating comprehensive information into well-informed decisions by fine-tuning them.

\subsubsection{Layered long-term memory}
\label{layered_long}

\textsc{FinMem}'s long-term memory organizes hierarchical financial data insights in a stratified structure, as illustrated in the lower section of Figure \ref{fig1:training_test_logic}. Drawing inspiration from the varying decay speeds in the human cognitive system's information processing layers \cite{craik1972levels}, \textsc{FinMem} employs a layered structure to accommodate the diverse time sensitivities inherent to different types of financial data. This structure categorizes summarized insights by their timeliness and decay rates. Insights are derived by the working memory's summarization operation. Those directed to deeper layers receive smaller decay rates, indicating longer retention, while those in shallower layers are assigned larger decay rates for shorter retention.

\begin{equation}
\gamma_l^{E} =   S_{\text{Recency}_{l}}^{E} + S_{\text{Relevancy}_{l}}^{E} + S_{\text{Importance}_{l}}^{E},
 \label{eqn:eq4}
\end{equation}
where each memory event is only associated with one score and can only belong to a single layer.
\\

Upon receiving an investment inquiry, \textsc{FinMem} retrieves the top-$K$ pivotal memory events from each layer and channels them to the immediate reflection component of the working memory. These events are chosen according to the descending order of their information retrieval score, denoted as $\gamma_l^{E}$, where $l$ belongs to the set {shallow, intermediate, deep}, as specified in Equation~\ref{eqn:eq4}. $E$ denotes a given memory event. This score, adapted from Park et al. \cite{10.1145/3586183.3606763} but with modified recency and importance computations, especially tailoring to handle data with various timelines. It encapsulates three metrics: recency, relevancy, and importance. Individual metric scores exceeding 1.0 are scaled to the [0,1] range before being summed. The modification is to achieve the layered processing function and represent the various periodicity of the financial environment. 
\begin{equation}
  \begin{split}
& S_{\text{Recency}_{l}}^{E} = e^{-\frac{\delta^{E}}{Q_l}},  \quad \; \delta^{E} = t_{\text{P}} - t_{E}, \\
 \end{split}
 \label{eqn:eq1}
  \end{equation}

where $\delta^{E}$ refers to the time difference between the memory event occurrence and the trading inquiry arrival. $Q_{\text{shallow}} = 14$, $Q_{\text{intermediate}} = 90$, and $Q_{\text{deep}} = 365$ correspond to day counts of two weeks, a quarter, and a year for shallow, intermediate, and deep processing layers, respectively.
\\

Upon a trade inquiry $P$ arrival in processing layer $l$ via LLM prompt, the agent computes the recency score $S_{\text{Recency}_{l}}^{E}$ per Equation.\ref{eqn:eq1}. $S_{\text{Recency}_{l}}^{E}$  inversely correlates with the time gap between the inquiry and the event's memory timestamp, mirroring Ebbinghaus's forgetting curve \cite{murre2015replication}. The stability term $Q_l$ in Equation.\ref{eqn:eq1} partially controls memory decay rates across layers, indicating longer memory persistence in the long-term layer with a higher stability value. In the context of trading, company annual reports, such as Form 10-Ks, are considered to have more extended timeliness compared to daily financial news. Therefore, they are assigned a higher stability value and are categorized within the deeper processing layer. This classification reflects their extended relevance and impact in financial decision-making scenarios.

\begin{equation}
  \begin{split}
& S_{\text{Relevancy}_{l}}^{E} = \frac{\mathbf{m_{E}} \cdot \mathbf{m_{P}}}{\|\mathbf{m_{E}}\|_2 \times \|\mathbf{m_{P}}\|_2}
 \end{split}
 \label{eqn:eq2}
  \end{equation}

The relevancy score, denoted as $S_{\text{relevancy}_{l}}^{E}$, quantifies the cosine similarity between the embedding vectors. These vectors are derived from the textual content of the memory event, $\mathbf{m_{E}}$, and the LLM prompt query, $\mathbf{m_{P}}$, using OpenAI's “text-embedding-ada-002” model, as depicted in Equation \ref{eqn:eq2}. The LLM prompt query incorporates inputs related to trading inquiries and the trading agent's character setting.

The importance score $S_{\text{Importance}_{l}}^{E}$ is computed using the value $v_l^{E}$ from a uniform piecewise scoring function (Formula~\ref{eqn:eq3}), multiplied by a degrading ratio $\theta_l$ (Formula~\ref{eqn:eq4}) as per Equation~\ref{eqn:eq5}. The likelihood of higher $v_l^{E}$ values increases from shallow to deep layers. $\theta_l$ measures the diminishing importance of an event over time, which has a close form design of \cite{10.1145/3586183.3606763}. But our approach tailors $\theta_l$ to the stratified structure of long-term memory. It adopts unique exponential functions for each layer. The base $\alpha_l$ for each layer is a hyperparameter, set to follow the sequence: $\alpha_{shallow} < \alpha_{intermediate} < \alpha_{deep}$. These values correlate with the rate at which their importance degrades after a certain period, providing another angle to measure importance variances across different memory types.  Through experimentation, we set $\alpha_{shallow} = 0.9$, $\alpha_{intermediate} = 0.967$ and $\alpha_{deep} = 0.988$. This ensures that $\theta_l$ decreases to a threshold score of $5$ after intervals of $30, 90,$ and $365$ days for shallow, intermediate, and deep layers, respectively.  The three-piece-wise functions for $S_{\text{Importance}_{l}}^{E}$ and $S_{\text{Recency}_{l}}^{E}$ enable \textsc{FinMem} to have layered processing in the long-term memory component.  Memory events are purged when $S_{\text{Recency}_{l}}^{E}$ is below $0.05$ or $S_{\text{Importance}_{l}}^{E}$ is under $5$ (pre-scaling).

\begin{equation}
  \begin{split}
& v_{l}^{E} = \begin{cases} 
40 & \text{with probability } p_1\\
60 & \text{with probability } p_2\\
80 & \text{with probability } p_3
\end{cases}
 \end{split}
 \label{eqn:eq3}
  \end{equation}


\begin{equation}
  \begin{split}
    \theta_{l} = (\alpha_l)^{\delta^{E}}, \quad & l = \text{shallow}, \text{intermediate}, \text{deep},
  \end{split}
  \label{eqn:eq4}
\end{equation}

where $p_1 + p_2 + p_3 = 1$, but their values vary by shallow, intermediate, and deep processing. when shallow processing ${p_1, p_2, p_3} = \{0.8, 0.15, 0.05\}$, intermediate processing, ${p_1, p_2, p_3} = \{0.05, 0.8, 0.15\}$ and deep processing, ${p_1, p_2, p_3} = \{0.05, 0.15, 0.8\}$. 

\begin{equation}
S_{\text{Importance}_{l}}^{E} = v_{l}^{E} * \theta_{l},
\label{eqn:eq5}
\end{equation}

Furthermore, an access counter function oversees the transfer of memory events among layers, ensuring that significant events influencing trading decisions ascend from shallower to deeper layers for extended retention and recurrent access by \textsc{FinMem}. Conversely, less pertinent events gradually diminish. This process is facilitated by the LLM validation tool Guardrails AI \cite{guardrailsai}, which monitors critical memory IDs across different layers. An event identified as pivotal for investment success receives an additional $5$ points in its importance score $S_{\text{Importance}_{l}}^{E}$. Upon meeting the criteria for upgrading to a deeper layer, an event's recency score $S_{\text{Recency}_{l}}^{E}$ is reset to $1.0$, emphasizing its importance and preventing rapid decay. By implementing this access counter, \textsc{FinMem} effectively identifies and prioritizes key events, taking into account their nature and frequency of retrieval.

\subsection{Decision-making Module}
\label{mem_form}

The decision-making module of \textsc{FinMem} efficiently integrates operational outcomes from the profiling and memory modules to support well-informed investment decisions, as depicted in Figure~\ref{fig2:workflow} (1). In its daily trading decisions, \textsc{FinMem} is asked to select from three distinct actions for a single share of a specific stock by Guardrails AI text validation function: “Buy”, “Sell”, or “Hold”. Additionally, the inputs and results required by \textsc{FinMem}'s decision-making module vary between its training and testing phases, with each phase's specifics detailed as follows:

During the training phase, \textsc{FinMem} accesses a wide array of multi-source information relevant to the entire time period. When \textsc{FinMem} is prompted with trading inquiries containing stock ticker and date, as well as trader character-related texts, it concurrently initiates observation and summarization operations in its working memory. \textsc{FinMem} observes the market ground labels mentioned in the description about the observation operation in Section~\ref{working memory}, which involve daily adjusted price differences between consecutive days, indicative of “Buy” or “Sell” actions. Utilizing these price change signals, \textsc{FinMem} identifies and prioritizes the top-$K$ memories, ranking them based on retrieval scores from each long-term memory layer. This procedure enables \textsc{FinMem} to produce comprehensive reflections that provide a well-founded rationale and in-depth inference of the correlation between market ground labels and the memories retrieved. Through repeated trading operations, reflections, and memory events with significant impact, transition to a deeper memory processing layer, getting preserved for guiding future investment decisions during the testing phase. 

In the testing phase, where \textsc{FinMem} cannot access future price data, it relies on the Cumulative Return over the previous $M$ trading days to anticipate future market trends. To compensate for the absence of future market price information, \textsc{FinMem} utilizes enhanced reflections derived from immediate reflections spanning an $M$-trading-day period as supplementary references. When faced with a specific trading inquiry, \textsc{FinMem} integrates insights from various sources, including historical Cumulative Return, outcomes from extended reflection, and the Top-$K$ retrieved memories. This comprehensive approach enables \textsc{FinMem} to execute well-informed trading decisions.

It should be noted that \textsc{FinMem} generates executable actions exclusively in the immediate reflection operation of the testing phase. Since the trading direction is guided by the actual price trend, the training phase of \textsc{FinMem} does not make investment decisions. Instead, this phase is dedicated to accumulating trading experience through comparing market trends with incoming multi-source financial messages. Additionally, during this phase, \textsc{FinMem} develops a memory module enriched with a comprehensive knowledge base, thereby evolving its capability for independent decision-making in future trading activities.

\section{Experiments Setups}
We aim to evaluate the trading performance of \textsc{FinMem}. And we further illustrate its unique advantages of requiring significantly less historical trading time window to train and take full use of key financial data time series as well as textual information. Specifically, we conducted several experiments to study the following research questions (\textbf{RQs}):

\begin{itemize}
\item \textbf{RQ1:} Is \textsc{FinMem} capable of outperforming contemporary state-of-the-art algorithmic trading agents?
\item \textbf{RQ2:} Are there tasks that challenge other trading algorithms but are manageable by \textsc{FinMem}?
\item \textbf{RQ3:} Which LLM is best suited to form the backbone framework of \textsc{FinMem}?
\item \textbf{RQ4:} Does equipping \textsc{FinMem} with different risk inclination choices truly differentiate its trading performance?
\item \textbf{RQ5:} Can \textsc{FinMem} effectively filter and prioritize information to facilitate informed trading decisions?  
\end{itemize}

In the rest of the section, we begin by introducing the real-world financial dataset used in our experiments. We then describe the comparative algorithmic agents and list several widely used financial metrics. Our experiments fall into two categories: 1) The comparative experiments of \textsc{FinMem} versus other algorithmic trading agents, and \textsc{FinMem} using different LLMs as backbone algorithms. 2) The ablation studies evaluate the effects of \textsc{FinMem}'s adjustable cognitive span and the role of the trader's dynamic character settings, particularly the risk inclinations, on its trading performance. Through experiments, \textsc{FinMem} demonstrates to outperform other comparative algorithmic agents. Furthermore, we are able to show that its profiling and memory modules are sophisticated and tailored to effectively address the intricacies of the financial landscape, resulting in superior trading performance.

\subsection{Datasets And Database Structure:} 

\begin{figure} 
    \centering
    \includegraphics[width=0.6\linewidth]{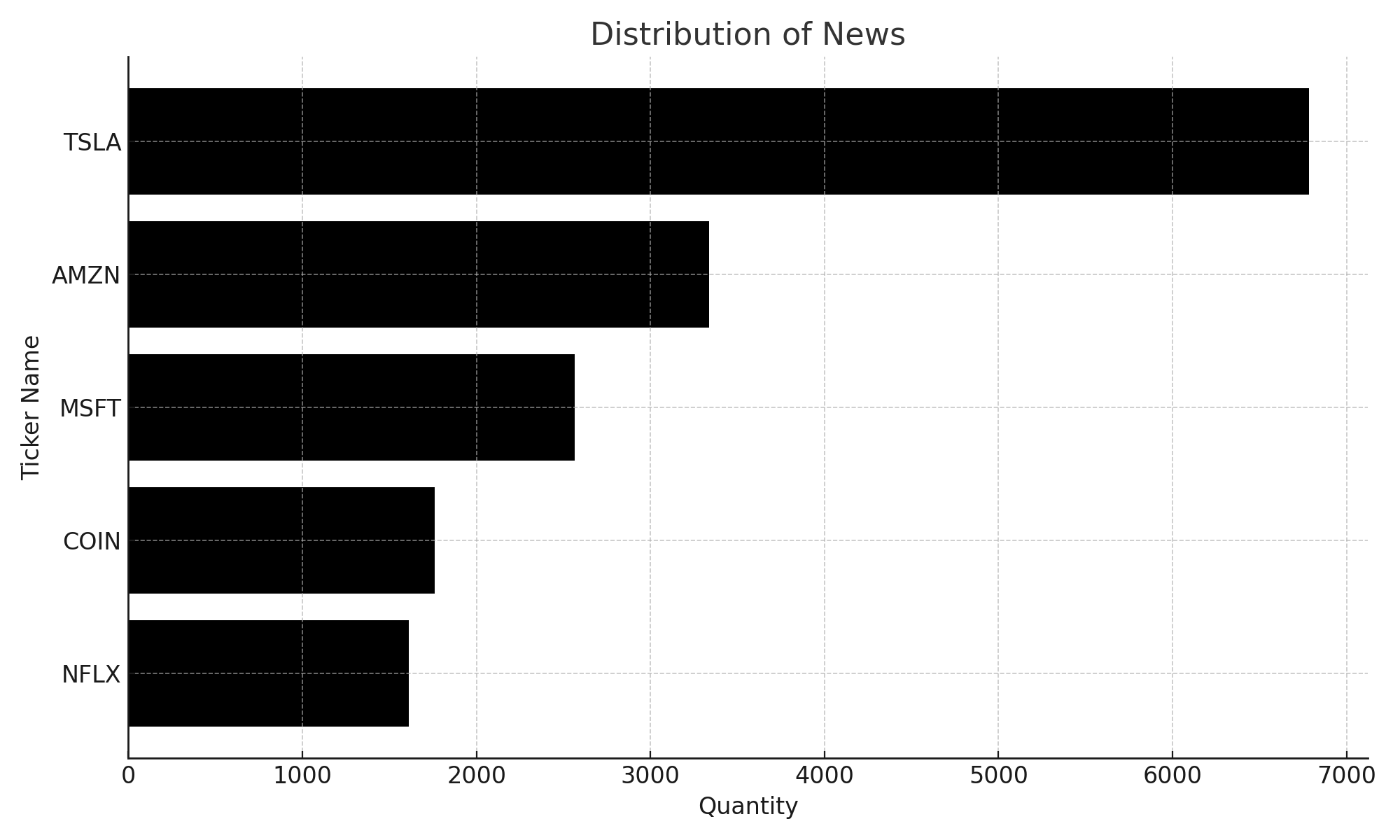}
  \caption{The distribution of news in scraped from Alpaca News API for the five stocks in the experiments}
  \label{fig:news_counts} 
\end{figure}

We assessed \textsc{FinMem}'s performance using multi-source financial data from August 15, 2021, to April 25, 2023, sourced from reputable financial databases and APIs like Yahoo Finance (via yfinance) and Alpaca News API, detailed explained in Table~\ref{tab:data_summary}. The stock tickers used in our comparative experiments are detailed in Figure~\ref{fig:news_counts}. These were selected because they are among those with the highest volumes of accessible news text data, and they are spread across various trading sectors.  This selection provides ample data to evaluate \textsc{FinMem}'s generalization capabilities. Additionally, Tesla, Inc. (TSLA) was specifically chosen for ablation studies due to its association with the largest amount of textual data, offering sufficient information to assess performance differences for the \textsc{FinMem}'s key features like cognitive spans.

The raw multi-source input data, initially stored in the “Raw Financial Data Warehouse”, are diverged into \textsc{FinMem}’s “Layered Long-term Memory Data Warehouse” based on timeliness through working memory's summarization operation in Figure~\ref{fig1:training_test_logic}. The deep processing layer holds annual reports (Form 10K's) insights, the intermediate layer contains quarterly reports (Form 10Q's) insights, and the shallow layer accommodates daily financial news insights.

We leveraged the open-source vector database FAISS \cite{johnson2019billion} for constructing the memory warehouse of \textsc{FinMem}, benefiting from its rapid querying in high-dimensional vectors and compatibility with OpenAI for cosine-similarity-based semantic searches on specific tickers. This setup facilitates efficient top-ranked event retrieval. Data categorization and memory module workflow are also illustrated in Figure~\ref{fig1:training_test_logic}

\label{data_warehouse}
\begin{table}[h]
    \centering
    \renewcommand{\arraystretch}{1.35}
    \begin{tabular}{|m{16.2cm}|}
    \hline
    \textbf{Raw Financial Data Warehouse} \\ 
    \hline
    \textbf{\textit{News data associated with ticker indexes:}} News data is sourced from the Alpaca News API, which utilizes Benzinga as its backend provider.\\ 
    \hline
    
    \textbf{\textit{Corporate quarter filings indexes:}}  Quarterly reports (Form 10-Q) are required by the U.S. Securities and Exchange Commission (SEC). \\ 
    \hline
    
     \textbf{\textit{Corporate annual filings indexes:}}  Annual reports (Form 10-K)
     required by the U.S. Securities and Exchange Commission (SEC).  \\ 
    \hline

    \textbf{\textit{Stock price records:}} Daily stock open-high-close-volume (OHLCV) data from Yahoo Finance.\\ 
    \hline
    \midrule
    \hline
    \textbf{\textsc{FinMem}'s Layered Long-term Memory Data Warehouse} \\ 
    \hline
    \textbf{\textit{Shallow Processing:}} Insights of real-time market news extracted by LLM. Updated daily.\\ 
    \hline
    
    \textbf{\textit{Intermediate Processing:}} Insights of 10-Q filings extracted by LLM. Updated quarterly.\\ 
    \hline
    
    \textbf{\textit{Deep Processing:}} 10-K fillings, all summarized by LLM. Updated yearly. Extended reflections for the stock in response to the trading inquiry include \textsc{FinMem}'s cumulative trading returns, decision-making processes, trade volumes, and underlying reasons. Updated daily. \\ 
    \hline
    \end{tabular}
    \caption{Raw data and memory warehouses of \textsc{FinMem}}
    \label{tab:data_summary}
\end{table}

\subsection{Baseline And Comparative models:} 

We assess \textsc{FinMem}'s trading performance in comparison to five advanced algorithmic agents and a commonly accepted baseline trading strategy. Among these, three models employ Deep Reinforcement Learning (DRL) approaches, while the remaining two are based on Large Language LLMs. Brief descriptions of each are provided below:

\textbf{Buy-and-Hold strategy (B\&H)}: 

A passive investment approach, where an investor purchases stocks and holds onto them for an extended period regardless of market fluctuations, is commonly used as a baseline for comparison of stock trading strategies. 

\textbf{DRL trading agents:} 

As the \textsc{FinMem} is practiced and examined on the basis of single stock trading and discrete trading actions, we choose three advanced DRL algorithms fitting into the same scenarios according to the previous and shown expressive performance in the work of Liu et al.  \cite{liu2021finrl, liu2022finrl}. The DRL training agents only take numeric features as inputs.

\begin{itemize}

\item \textbf{Proximal Policy Optimization (\textsc{PPO}):} \textsc{PPO} \cite{schulman2017proximal} is employed in stock trading due to its stability and efficiency. One salient advantage of PPO is that it maintains a balance between exploration and exploitation by bounding the policy update, preventing drastic policy changes. 
\item \textbf{Deep Q-Network (\textsc{DQN}):} \textsc{DQN} \cite{mnih2013playing} is an adaptation of Q-learning, that can be used to optimize investment strategies. Unlike traditional Q-learning that relies on a tabular approach for storing Q-values, DQN generalizes Q-value estimation across states using deep learning, making it more scalable for complex trading environments. 
\item \textbf{Advantage Actor-Critic (\textsc{A2C}):} \textsc{A2C} \cite{mnih2016asynchronous} is applied to optimize trading actions in the financial environment. It operates by simultaneously updating both the policy (actor) and the value (critic) functions, providing a balance between exploration and exploitation. 

\end{itemize}
\textbf{LLM trading agents:} 

We evaluate \textsc{FinMem} against two LLM agents in the context of stock trading. The first LLM agent, known for its proficiency in general-purpose tasks, serves as a baseline. The second agent, a leading-edge LLM in trading, has been acclaimed for its promising performance in stock market operations.

\begin{itemize}
\item \textbf{General-purpose Generative Agents -- \textsc{GA}:} The generative AI agent by Park et al. \cite{park2022social}, originally intended to simulate realistic human behavior and make everyday decisions, has been adapted here for specific stock trading tasks. This agent's architecture includes a memory module that employs recency, relevance, and importance metrics to extract pivotal memory events for informed decision-making. However, it does not provide a layered memory module to effectively differentiate the time sensitivities unique to various types of financial data. Additionally, although it features a profiling module to define agent attributes like professional background, the model does not include a mechanism for self-adaptive risk preference. In our experiments, we modified the original prompt template created by Park et al., which was intended for general daily tasks, to suit financial investment tasks. The textual elements of this revised template closely align with those of \textsc{FinMem}, with the exception of two components that are absent in this version of general-purpose Generative Agents.

\item \textbf{LLM trading agents -- \textsc{FinGPT}:} A novel open-source LLM framework specialized for converting incoming textual and numeric information into informed financial decision-making, introduced by Yang et al. \cite{yang2023fingpt}. It claims superiority over the traditional buy-and-hold strategy.
\end{itemize}

\subsection{Evaluation Metrics:}

\begin{figure*} 
    \centering
    \includegraphics[width=\linewidth]{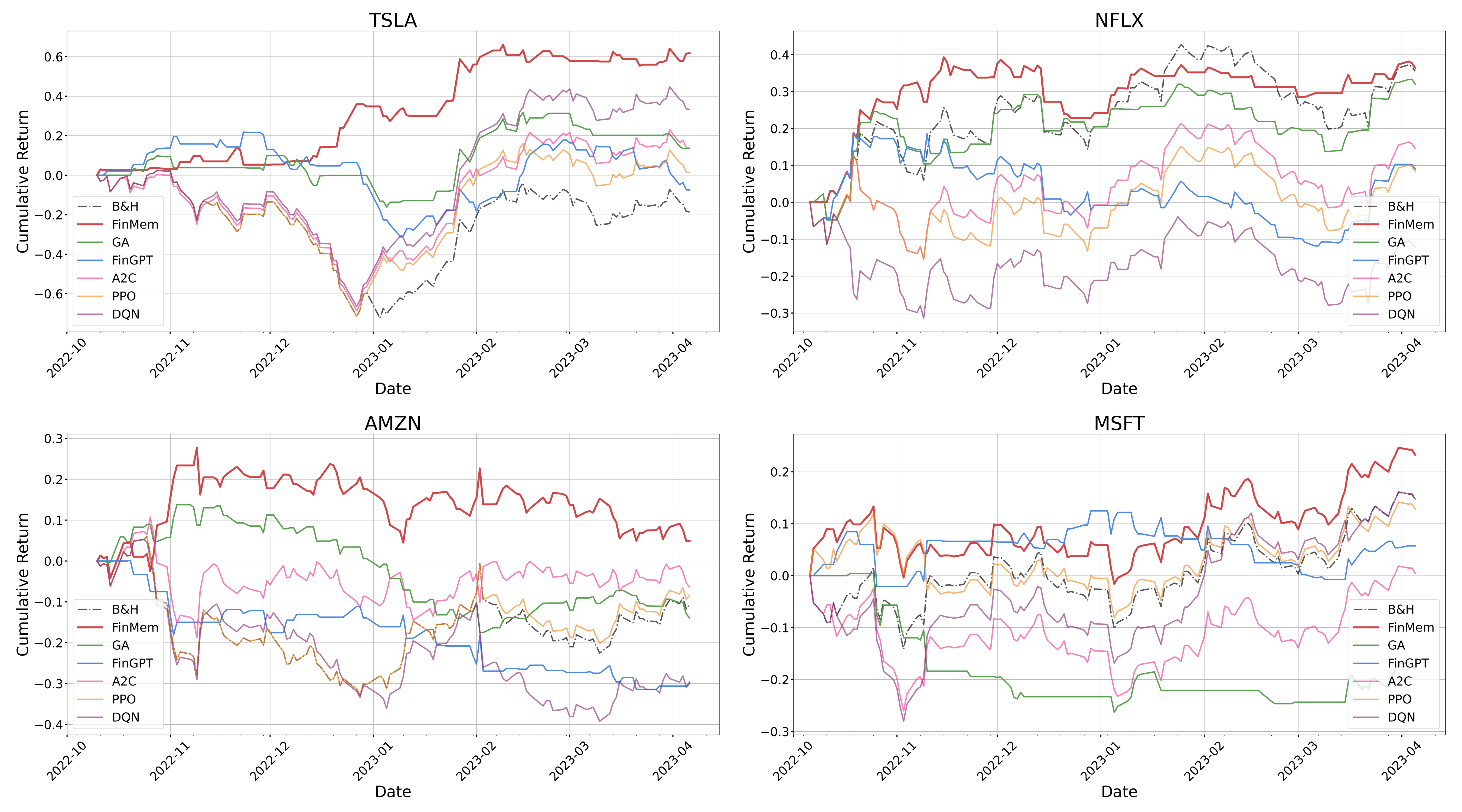}
  \caption{Cumulative return comparison over time between \textsc{FinMem} and other algorithmic agents across five stocks.}
  \label{fig1:four_stock_comparison} 
\end{figure*}

\begin{figure} 
    \centering
    \includegraphics[width=0.5\linewidth]{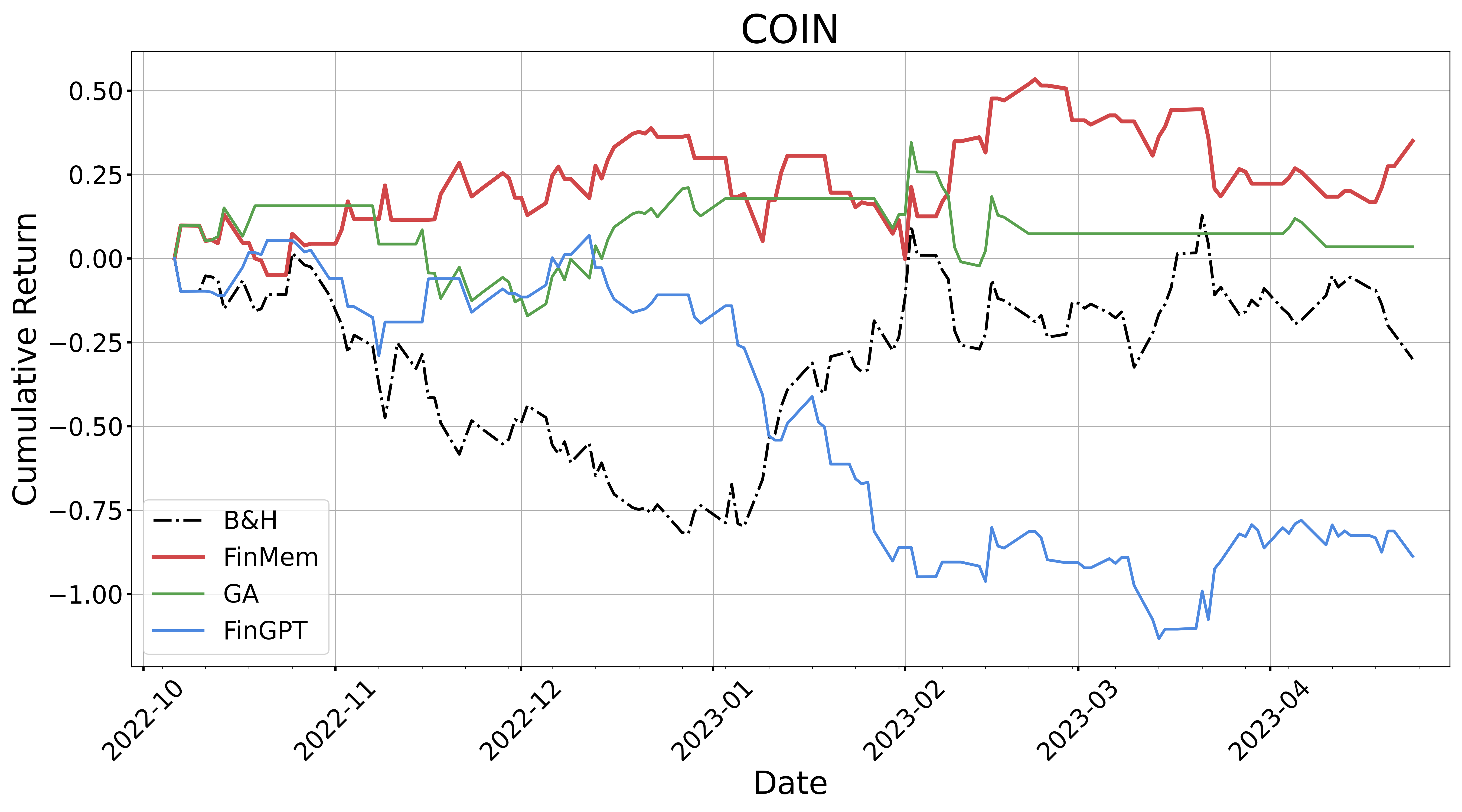}
  \caption{Cumulative Return comparison over time between \textsc{FinMem} and other algorithmic agents on Coinbase Global, Inc. (COIN).}
  \label{fig:coin} 
\end{figure}

We employ five widely-used metrics in finance to compare the investment rewards of \textsc{FinMem} against other algorithmic trading agents. Here are their introductions: 
\begin{itemize}

  \item \textbf{Cumulative Return} \cite{hull2007risk}: Cumulative Return is a key trading performance metric because it provides a comprehensive insight into investment performance, especially for strategies that emphasize long-term growth and reinvestment. The effectiveness of different investment strategies is evaluated based on their Cumulative Returns, which reflect the total change in value over time. In this study, we compute Cumulative Returns over the specified period by summing daily logarithmic returns, as outlined in Equation~\ref{eq:cum_return}. This method is widely accepted in the finance area due to its ability to precisely capture minor price fluctuations and symmetrically address gains and losses. In essence, a higher Cumulative Return typically indicates a more effective strategy.

\begin{align}
    \textbf{Cumulative Return} &= \sum_{t=1}^{n} r_i \nonumber \\
                                   &= \sum_{t=1}^{n} \left[ \ln\left(\frac{p_{t+1}}{p_t}\right) \cdot \text{action}_t \right],
    \label{eq:cum_return}
\end{align}

where $r_i$ represents the logarithmic return for day $t+1$, $p_t$ is the closing price on day $t$, $p_{t+1}$ is the closing price on day $t+1$, and $\text{action}_t$ denotes the trading decision made by the model for that day.

  \item \textbf{Sharpe Ratio} \cite{sharpe1994sharpe}: Sharpe Ratio is another core metric for evaluating investment performance and adjusting returns for risk. It is calculated by dividing the portfolio's average excess return ($R_p$) over the risk-free rate ($R_f$) by its volatility ($\sigma_p$), as shown in Equation~\ref{eq:sharpe}. This metric adjusts returns for risk, with a higher ratio indicating better risk-adjusted performance. Essential in comparing different portfolios or strategies, it contextualizes performance against similar investments. Although a Sharpe Ratio above 1 is typically considered favorable and above 2 as excellent, these benchmarks can vary depending on the context of comparison.
  
  \begin{equation}
    \textbf{Sharpe Ratio} = \frac{R_p - R_f}{\sigma_p}
    \label{eq:sharpe}
  \end{equation}  
   

  \item  \textbf{Annualized Volatility and Daily Volatility}\cite{cochrane1988volatility}: Annualized Volatility (Annum-Volatility) is calculated as the \textbf{Daily Volatility} (standard deviation of daily logarithmic returns) multiplied by the square root of the typical number of trading days in a year (252) as outlined in Equation~\ref{eq:annuaVol}, is vital for assessing investment risk. This measure reflects the extent of fluctuation in a security or market index's returns over a year, indicating potential deviations from average returns. It's especially relevant for investors with specific risk profiles, such as those who are risk-averse, who may prefer portfolios demonstrating lower annualized volatility.  

   \begin{align}
   \label{eq:annuaVol}
    \textbf{Annum-Volatility} &= \textbf{Daily Volatility} \times \sqrt{252} 
   \end{align} 
  

\item \textbf{Max Drawdown} \cite{ang2003downside}: Max Drawdown is a metric for assessing risk. It represents the most significant decrease in a portfolio's value, from its highest ($P_{\text{peak}}$) to its lowest point ($P_{\text{trough}}$) until a new peak emerges, detailed in Equation~\ref{eq:maxdrawdown}. Indicative of investment strategy robustness, a smaller Max Drawdown suggests reduced risk.
  
    \begin{align}
    \label{eq:maxdrawdown}
    \textbf{Max Drawdown} = \text{max}(\frac{P_{\text{peak}} - P_{\text{trough}}}{P_{\text{peak}}})
    \end{align}
  
\end{itemize}

In our experiments and ablation studies, we recorded the metric outcomes as an average from five repeated trials.

\begin{table*}[htbp]
\centering
\begin{adjustbox}{max width=\dimexpr\columnwidth-0.1cm,center}
\begin{tabular}{llccccc}
\toprule
\textbf{Ticker} & \textbf{Model} & \textbf{Cumulative Return (\%)} & \textbf{Sharpe Ratio} & \textbf{Daily Volatility (\%)} & \textbf{Annualized Volatility (\%)} & \textbf{Max Drawdown (\%)} \\
\midrule
\multirow{7}{*}{TSLA} 
    & Buy and Hold & -18.6312 & -0.5410 & 4.4084 & 69.9818 & 55.3208 \\
    & \textsc{FinMem} & \textcolor{FinMem}{\textbf{61.7758*}} & \textbf{2.6789} & 2.9522 & 46.8649 & \textbf{10.7996} \\
    & Generative Agents & 13.4636 & 0.5990 & \textbf{2.8774} & \textbf{45.6774} & 24.3177 \\
    & FinGPT & -7.4554 & -0.2795 & 3.4145 & 54.2027 & 42.3993 \\
    & A2C & 13.7067 & 0.3979 & 4.4096 & 70.0009 & 52.3308 \\
    & PPO & 1.2877 & 0.0374 & 4.4110 & 70.0232 & 54.3264 \\
    & DQN & \textcolor{Park}{33.3393} & 0.9694 & 4.4027 & 69.8900 & 52.0033 \\
\midrule
\multirow{7}{*}{NFLX} 
    & Buy and Hold & 35.5111 & 1.4109 & 3.1964 & 50.7410 & 20.9263 \\
    & \textsc{FinMem} & \textcolor{FinMem}{\textbf{36.4485*}} & \textbf{2.0168} & \textbf{2.2951} & \textbf{36.4342} & \textbf{15.8495} \\
    & Generative Agents & \textcolor{Park}{32.0058} & 1.5965 & 2.5460 & 40.4168 & 16.9893 \\
    & FinGPT & 9.0090 & 0.4266 & 2.6819 & 42.5732 & 28.2705 \\
    & A2C & 14.6155 & 0.5788 & 3.2071 & 50.9112 & 25.0184 \\
    & PPO & 8.4121 & 0.3330 & 3.2086 & 50.9344 & 25.0184 \\
    & DQN & -12.2067 & -0.4833 & 3.2078 & 50.9217 & 28.7017 \\
\midrule
\multirow{7}{*}{AMZN} 
    & Buy and Hold & -10.7739 & -0.4980 & 2.7697 & 43.9674 & 33.6828 \\
    & \textsc{FinMem} & \textcolor{FinMem}{\textbf{4.8850*}} & \textbf{0.2327} & 2.6872 & 42.6576 & \textbf{22.9294} \\
    & Generative Agents & -13.9271 & -0.9981 & 1.7864 & 28.3576 & 27.7334 \\
    & FinGPT & -29.6781 & -2.1756 & \textbf{1.7464} & \textbf{27.7225} & 28.4838 \\
    & A2C & \textcolor{Park}{-6.3591} & -0.2938 & 2.7706 & 43.9819 & 26.1275 \\
    & PPO & -8.4194 & -0.3891 & 2.7702 & 43.9761 & 33.6828 \\
    & DQN & -29.9820 & -1.3906 & 2.7603 & 43.8177 & 38.3740 \\
\midrule
\multirow{7}{*}{MSFT} 
    & Buy and Hold & 14.6949 & 0.8359 & 2.2326 & 35.4411 & 15.0097 \\
    & \textsc{FinMem} & \textcolor{FinMem}{\textbf{23.2613*}} & \textbf{1.4402} & 2.0512 & 32.5617 & 14.9889 \\
    & Generative Agents & -18.1031 & -1.6057 & \textbf{1.4318} & \textbf{22.7285} & 24.2074 \\
    & FinGPT & 5.7356 & 0.4430 & 1.6442 & 26.1008 & \textbf{12.8459} \\
    & A2C & 0.4598 & 0.0261 & 2.2357 & 35.4913 & 23.6781 \\
    & PPO & 12.8067 & 0.7282 & 2.2333 & 35.4532 & 19.5355 \\
    & DQN & \textcolor{Park}{14.7397} & 0.8385 & 2.2326 & 35.4408 & 25.1845 \\
\midrule
\multirow{7}{*}{COIN} 
    & Buy and Hold & -30.0071 & -0.5150 & 6.7517 & 107.1795 & 60.5084 \\
    & \textsc{FinMem} & \textcolor{FinMem}{\textbf{34.9832*}} & \textbf{0.7170} & 5.6538 & 89.7515 & 35.7526  \\
    & Generative Agents & \textcolor{Park}{3.4627} & 0.0896 & \textbf{4.4783} & \textbf{71.0908} & \textbf{32.0957} \\
    & FinGPT & -88.7805 & -1.9507 & 5.2736 & 83.7153 & 73.5774 \\
    & A2C & - & - & - & - & - \\
    & PPO & - & - & - & - & - \\
    & DQN & - & - & - & - & - \\
\bottomrule
\end{tabular}
\end{adjustbox}
\caption{Overall trading performance comparison during testing period between \textsc{FinMem} and other algorithmic agents across five stocks.\textit{* indicates that the result of the Wilcoxon signed-rank test is statistically significant.}\protect\footnotemark[3]}
\label{exp1:performance_overview}
\end{table*}

\begin{figure} 
    \includegraphics[width=\linewidth]{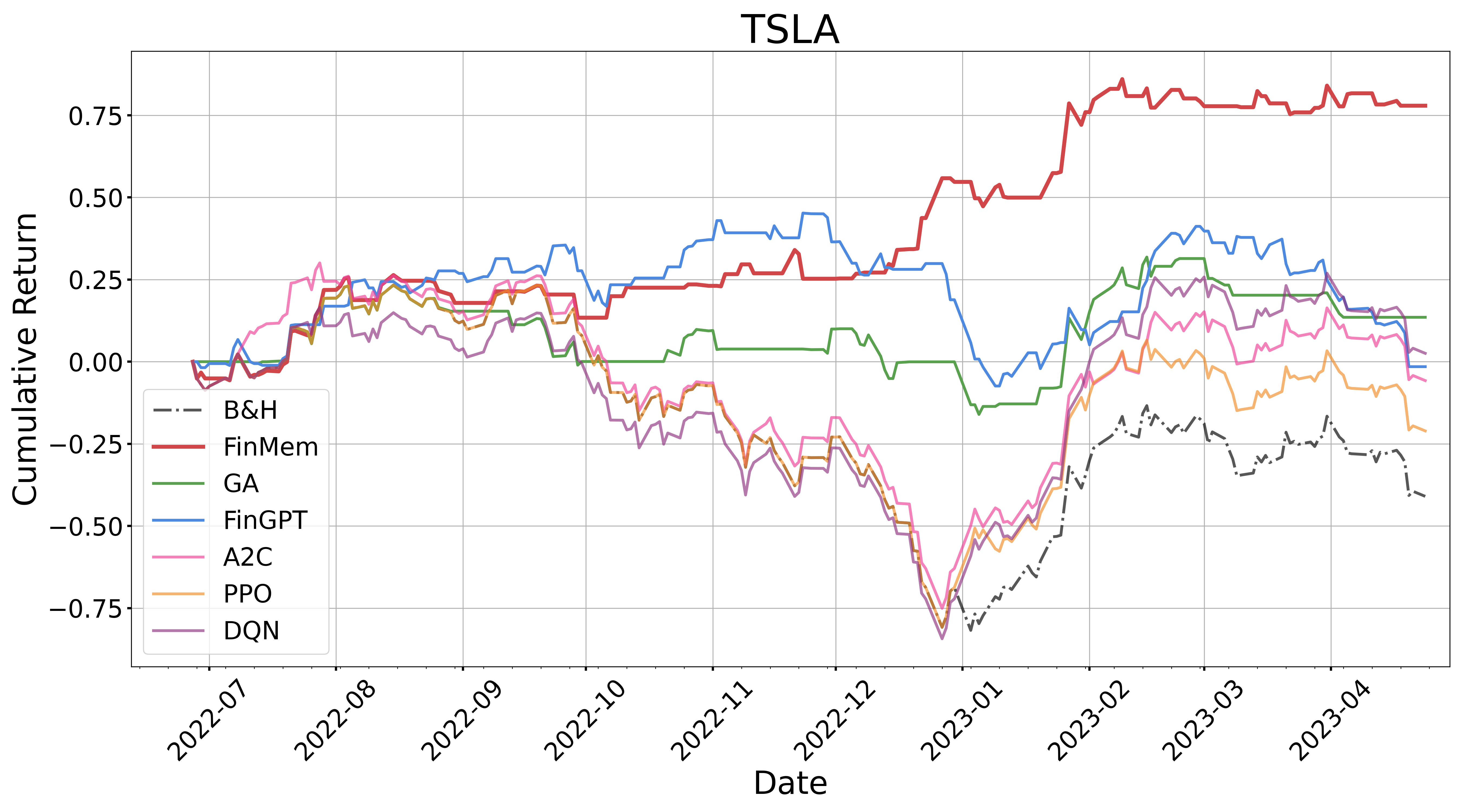}
  \caption{Cumulative Return of \textsc{FinMem} on trading Tesla, Inc. (TSLA) stock Over an Extended Testing Period.}
  \label{fig:tsla_full} 
\end{figure}

\section{Experiments:}
\label{big_exp}

\subsection{Implementation Details:}
\label{param_settings}

\footnotetext[3]{The bold numbers in this and subsequent tables signify the best performance for the respective metrics.}

In the Trading Agents Comparison, \textsc{FinMem} employs GPT-4-Turbo as its backbone algorithm. The temperature parameter of the model was set at 0.7 to maintain a balance between response content consistency and model creativity. It was trained on financial data from August 17, 2021, to October 05, 2022, and underwent testing with data from October 06, 2022, to April 10, 2023. The training period was chosen to account for the seasonal nature of corporate financial reporting and the duration of data retention in \textsc{FinMem}'s memory module. The selected training duration ensures the inclusion of at least one publication cycle of either Form 10-Q, classified as intermediate memory, or Form 10-K, regarded as deep memory, or in some instances, both. This strategy ensures that the experiences retained in \textsc{FinMem} are still influential during the testing phase for a significant period. Additionally, the training duration allowed \textsc{FinMem} sufficient time to establish inferential links between financial news, market indicators, and stock market trends, thereby accumulating substantial experience. Furthermore, we set the number of top memory events retrieved from each layer of long-term memory at 5. We ran \textsc{FinMem} using each of the three available risk inclination settings. The reported performance outcomes are based on the setting that achieved the highest cumulative return during the testing phase.

To maintain consistency in the comparison, the training and testing phases for the other two LLM-based agents were aligned with those of \textsc{FinMem}. For parameters of other LLM-based agents that are not encompassed by \textsc{FinMem}'s configuration, they were kept in accordance with their original settings as specified in their respective source codes.

Considering that DRL algorithms need extensive training data for stable and converged results, and given our daily evaluation of trading performance, we extended the DRL agents' training period to roughly a 10-year span, from January 1, 2012, to October 05, 2022, for a fair comparison. The testing period was kept consistent with the other models. The DRL algorithms were implemented using Stable Baselines 3 \cite{stable-baselines3}.


\textsc{FinMem}'s performance was benchmarked against that of the most effective comparative model, using Cumulative Return and Sharpe Ratio as the primary evaluation metrics. The statistical significance of \textsc{FinMem}'s superior performance was ascertained through the non-parametric Wilcoxon signed-rank test, which is particularly apt for the non-Gaussian distributed data.

\subsection{Algorithmic Trading Agents Comparison (RQ1 \& RQ2)}
In this experiment, we assess the stock trading performance of \textsc{FinMem} against other models, focusing on stocks from five companies in different trading sectors: Tesla, Inc. (TSLA), Netflix, Inc. (NFLX), Amazon.com, Inc. (AMZN), Microsoft Corporation (MSFT), and Coinbase Global, Inc. (COIN). The performance of all algorithmic trading agents across five key metrics is consolidated in Table~\ref{exp1:performance_overview}. Given the pivotal role of Cumulative Return in evaluating trading performance over time, we present detailed time series plots in Figure~\ref{fig1:four_stock_comparison} and Figure~\ref{fig:coin}. It's important to note that the trading performance of \textsc{FinMem} for COIN was exclusively compared with LLM trading agents and the baseline. This is because Coinbase Global, Inc. completed its IPO in April 2021 and, as a result, had not accumulated enough trading data to facilitate stable outcomes with Deep Reinforcement Learning (DRL) algorithms. These plots illustrate the changes in Cumulative Return for each of the five companies throughout the testing phase, offering an in-depth comparison of performance.

In response to \textbf{RQ1}, the trading outcomes presented in Table~\ref{exp1:performance_overview} reveal that \textsc{FinMem} outperforms all other algorithmic trading agents and the B\&H baseline strategy in terms of Cumulative Return and Sharpe Ratio. \textsc{FinMem}'s superiority is statistically significant when compared to the second-best trading strategy. Specifically, for TSLA and NFLX, \textsc{FinMem}'s strategy achieves Sharpe Ratios exceeding $2.0$ and Cumulative Returns surpassing $0.35$ while maintaining the lowest Volatility and Max Drawdown. These indicators underscore \textsc{FinMem}'s ability to generate higher returns per unit of risk. In the case of MSFT and NFLX, \textsc{FinMem} also records a Sharpe Ratio above $1.0$ and a Cumulative Return over $0.2$, coupled with relatively low Volatility and Max Drawdown, demonstrating its impressive trading performance. For AMZN and COIN, \textsc{FinMem} consistently delivers positive Cumulative Returns and superior Sharpe Ratios, outperforming other strategies that yield negative values for these metrics. Additionally, its Volatility and Max Drawdown are on the lower end. Hence, these results collectively demonstrate \textsc{FinMem}'s robust trading performance across a diverse range of trading sectors. Specifically, \textsc{FinMem} exhibits superior performance compared to the two other LLM agents in our study, \textsc{FinGPT} and the general-purpose generative agent developed by Park et al. This underscores the effectiveness of \textsc{FinMem}'s unique profiling and memory structure, which are particularly tailored for LLM agents dealing with financial data, significantly enhancing their investment decision-making capabilities.


\begin{figure} 
    \centering
    \includegraphics[width=0.66\linewidth]{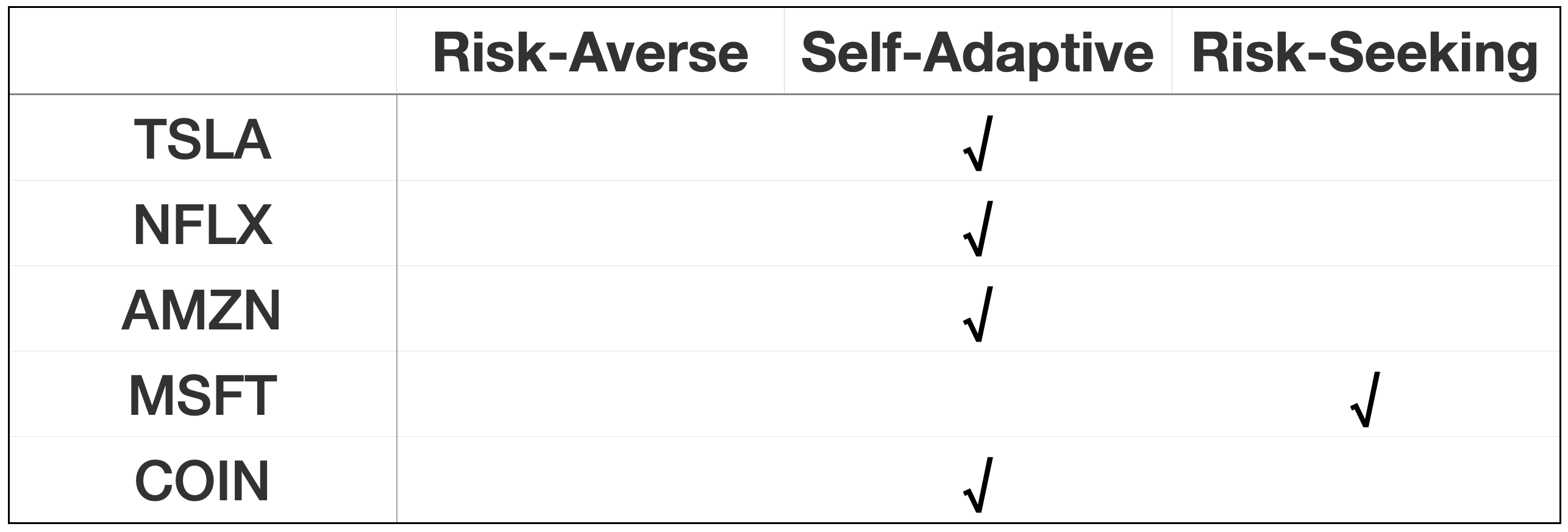}
  \caption{The optimal risk inclination for \textsc{FinMem} when trading different stocks.}
  \label{fig:character_options} 
\end{figure}

In response to \textbf{RQ2}, the main challenge for DRL trading agents is that they require training data with a large volume and extended time span, which are hard to achieve when operating on stocks with limited historical data. As shown in Table~\ref{exp1:performance_overview}, our experiments reveal that \textsc{FinMem} achieves superior trading performance with a much shorter training duration compared to DRL trading agents trained on data spanning nearly a decade. This efficiency makes \textsc{FinMem} particularly useful for newly public companies like Coinbase Global, Inc., which have limited trading histories. DRL agents often face convergence issues due to inadequate training data in such cases. Moreover, even among LLM-based trading agents suited for shorter training periods, \textsc{FinMem}'s performance stands out, as illustrated in Figure~\ref{fig:coin}.

To further assess \textsc{FinMem}'s adaptability to limited training data, we narrowed the training period down to an even shorter period, spanning from August 17, 2021, to February 10, 2022. We then extended the testing phase to cover from February 11, 2022, to April 25, 2023. This evaluation focused on the stock of Tesla, Inc., which has the largest volume of news data. The trading performance of \textsc{FinMem} during this period is depicted in Figure~\ref{fig:tsla_full}. Remarkably, using less than six months of daily frequency data for training, which encompassed the publication of one Form 10-K and one Form 10-Q, \textsc{FinMem} consistently ranked high in gains and attained the highest cumulative return after the latter half of December 2022.

The consistently strong trading performance of \textsc{FinMem} can be attributed to its innovative profiling and memory module design. This design enables \textsc{FinMem} to effectively integrate, comprehend, and prioritize key information from both textual and numerical data. The flexibility of \textsc{FinMem}'s profiling module in selecting risk inclinations plays a pivotal role in its ability to both exploit rising market trends and safeguard assets during downturns. A prime example is TSLA, which achieved its best trading results under a self-adaptive risk setting in \textsc{FinMem}. This configuration enables \textsc{FinMem} to pursue a conservative and cautious strategy when facing negative short-term cumulative returns. On the other hand, with positive short-term returns, \textsc{FinMem} switches to an optimistic and assertive approach, thus avoiding excessive passivity. This self-adaptive risk inclination proved effective for most stocks, apart from MSFT as shown in Table~\ref{fig:character_options}. For MSFT, a risk-seeking inclination was most beneficial, resonating with its general bullish in the stock market. Additionally, the memory module's core features, including varied retention times for different information types and critical memory transitions, equip \textsc{FinMem} to capture essential information for well-informed investment decisions.

\section{Ablation Studies}

We conducted three distinct ablation studies to evaluate key component alternatives in \textsc{FinMem}. These studies concentrated on the backbone algorithm, the memory module's cognitive capacity, and the character setting in the profiling module, particularly examining the aspect of risk inclination. These studies were done using the stock of Tesla, Inc., with a more compact training period from March 14, 2022, to June 15, 2022, and a testing period from June 16, 2022, to December 28, 2022. This shorter duration was chosen for budgetary efficiency, yet it remains sufficient to differentiate the functionality of each component.

\subsection{\textsc{FinMem} Backbone Algorithm Comparison (RQ3)}

\begin{table*}[htbp]
\centering
\begin{adjustbox}{max width=\dimexpr\columnwidth-0.1cm,center}
\begin{tabular}{lcccccc}
\toprule
\textbf{Metric} & \textbf{B\&H} & \textbf{GPT 3.5-Turbo} & \textbf{GPT4} & \textbf{GPT4-Turbo} & \textbf{davinci-003} & \textbf{Llama2-70b-chat} \\
\midrule
\textbf{Cumulative Return (\%)} & -66.9497 & 16.1501 & \textbf{62.6180} & 54.6958 & 1.6308 & -52.7233 \\
\midrule
\textbf{Sharpe Ratio} & -2.0845 & 2.1589 & 2.2251 & \textbf{2.4960} & 0.8515 & -2.8532 \\
\midrule
\textbf{Daily Volatility (\%)} & 3.8050 & 0.8862 & 3.3339 & 2.5960 & \textbf{0.2269} & 2.1891 \\
\midrule
\textbf{Annualized Volatility (\%)} & 60.4020 & 14.0683 & 52.9237 & 41.2100 & \textbf{3.6018} & 34.7503 \\
\midrule
\textbf{Max Drawdown (\%)} & 67.3269 & 1.1073 & 17.4012 & 12.5734 & \textbf{0.8408} & 44.7168 \\
\bottomrule
\end{tabular}
\end{adjustbox}
\caption{Comparison of trading performance during the testing period for \textsc{FinMem} using different LLMs as backbone algorithms.}
\label{tab:llm-backbones}
\end{table*}


In our first study, we evaluated the trading performance of \textsc{FinMem} using various LLMs as its backbone algorithms. The LLMs under consideration included davinci-003, GPT 3.5-Turbo, GPT4, GPT4-Turbo, and Llama2-70b-chat. The parameter settings were consistent with its optimal performance in the comparative experiment detailed in Section~\ref{big_exp}, and the risk inclination was configured to be self-adaptive. All other model settings were maintained as outlined in Section~\ref{param_settings}. The results of this evaluation are compiled in Table~\ref{tab:llm-backbones}.

\begin{figure} 
    \centering
    \includegraphics[width=0.88\linewidth]{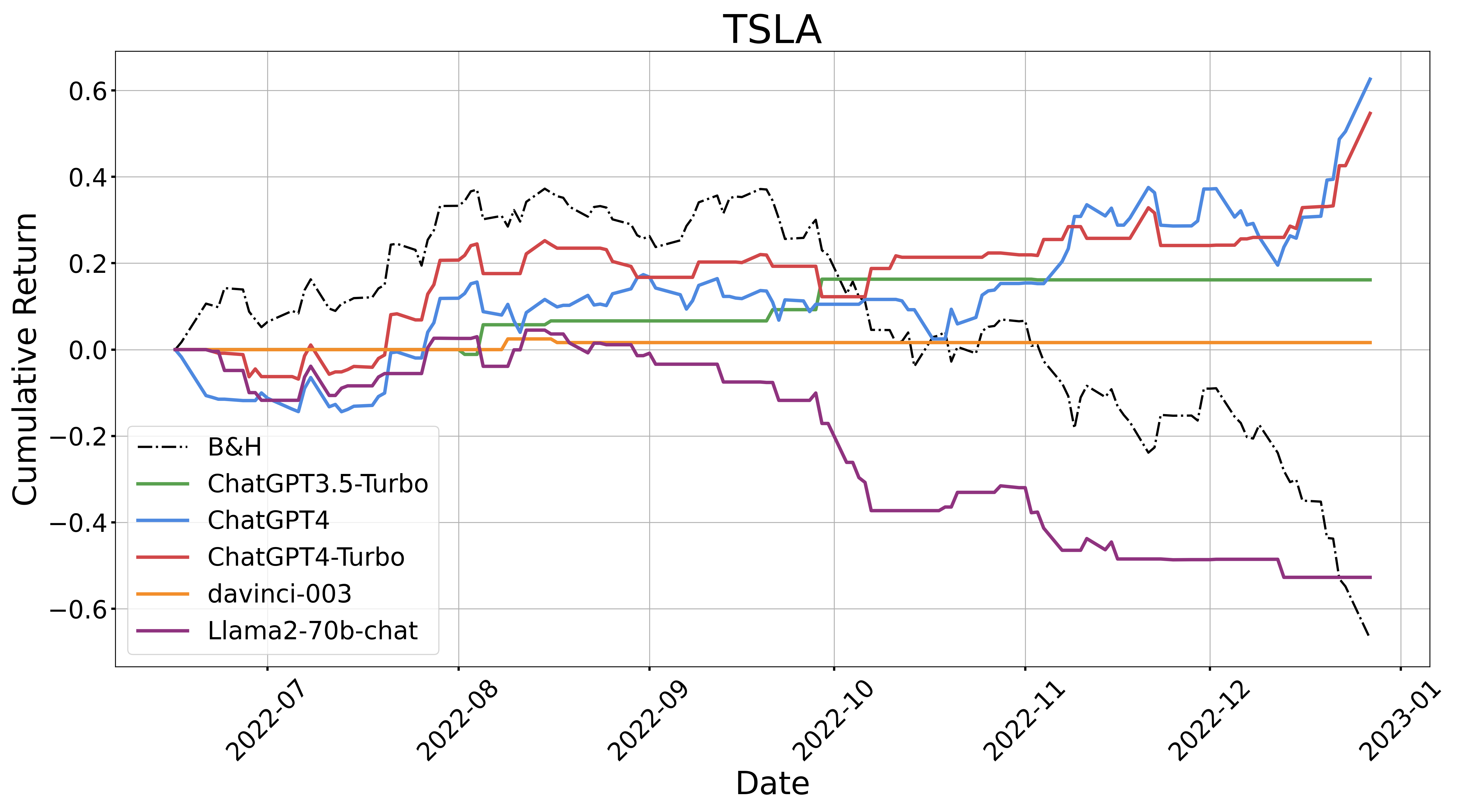}
  \caption{Comparison of overall Cumulative Returns over time for \textsc{FinMem} using different LLMs as backbone algorithms.}
  \label{fig:tsla_dif_LLM} 
\end{figure}

In response to \textbf{RQ3}, Addressing Research Question 3, the findings demonstrate that \textsc{FinMem}, powered by GPT-4 and GPT-4 Turbo, delivered superior trading results during the test phase. Specifically, GPT-4 recorded the highest cumulative return, while GPT-4-Turbo exhibited the most favorable Sharpe Ratio. GPT 3.5-Turbo's performance was also noteworthy, following closely behind. As depicted in Figure~\ref{fig:tsla_dif_LLM}, though slightly lower than market baseline (B\&H), \textsc{FinMem} with GPT-4-Turbo led in cumulative returns before October 2022. This period was characterized by relative stability and a modest upward trend in TSLA stock. After October 2022, with TSLA undergoing increased volatility and a notable downward trend, the cumulative return trajectory for \textsc{FinMem} with GPT-4-Turbo exhibited significantly lower volatility and sustained stable returns not markedly lower than those of GPT-4. These results indicate that GPT-4 Turbo is the most suitable backbone algorithm for \textsc{FinMem}.

\textsc{FinMem} configured with davinci-003 and Llama2-70b-chat exhibited the lowest Annualized Volatility and Max Drawdown, yet their Cumulative Return and Sharpe Ratio were underwhelming. As illustrated in Figure~\ref{fig:tsla_dif_LLM}, both models defaulted to a “Hold” strategy beyond a certain point during periods of intense fluctuation in TSLA stock. The unsatisfactory performance of davinci-003 may be attributed to its limited capability, as an earlier generation language model, to capture and understand nuanced yet decisive information.

We selected Llama2-70b-chat as it was deemed to possess stronger in-context learning and instruction-following capabilities compared to other Llama family models with fewer parameters, as noted in Zhao et al. \cite{zhao2023survey}. Nonetheless, in the context of stock trading, it still demonstrated challenges in adequately comprehending key messages necessary for effective trading decisions. The comparatively poorer performance of Llama2-70b-chat can also be attributed to its shorter context window, especially when compared to the GPT models. When integrated with \textsc{FinMem}, it needs to simplify prompts and shorten the length of retrieved memory insights, which could potentially result in some loss of context. The exceptional trading result demonstrated by GPT-4-Turbo across all models was a main factor in choosing it as the backbone algorithm for \textsc{FinMem} in our earlier comparative analysis with other algorithmic trading agents.

\subsection{Influence about varying the \textsc{FinMem} character design (RQ4)}

In our second study, we focused on evaluating the influence of \textsc{FinMem}'s profiling module on its trading effectiveness. Specifically, our assessment centered on the effects of customizing trader profiles according to specific stock trading, with a particular focus on risk inclination. As depicted in Figure~\ref{fig:character_options}, we equipped \textsc{FinMem} with three distinct risk profiles: risk-seeking, risk-averse, and a self-adaptive character. We executed a comparative analysis of \textsc{FinMem}'s performance across these risk profiles, maintaining consistency in all other settings as outlined in Section~\ref{param_settings}.

\begin{figure} 
    \centering
    \includegraphics[width=0.88\linewidth]{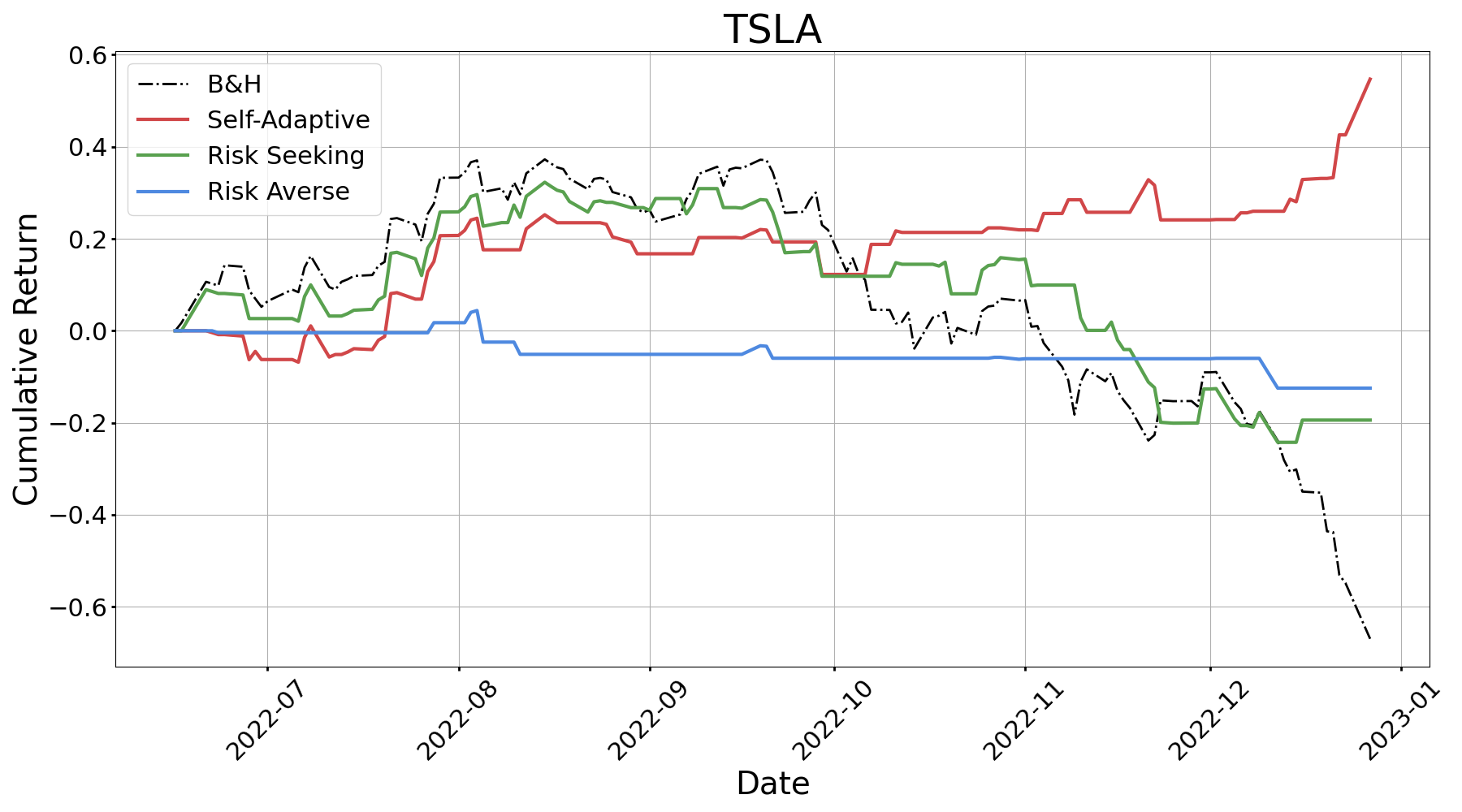}
  \caption{Comparison of Cumulative Return during with different risk inclinations setting in \textsc{FinMem}'s profiling module.}
  \label{fig6:diff_char} 
\end{figure}

In response to \textbf{RQ4}, Table~\ref{tab:diff_char_table} delineates the varied trading performance across different risk profiles. The self-adaptive profile enabled \textsc{FinMem} to achieve the most favorable trading performance, as it was the only one to secure a positive Cumulative Return and a Sharpe Ratio exceeding 2.0, along with the least Max Drawdown by the end of the testing phase. Figure~\ref{fig6:diff_char} illustrates \textsc{FinMem}'s capacity to adeptly navigate substantial stock price volatility and to strategically modulate its trading behavior when necessary. In contrast, the risk-seeking profile, while beneficial during a stable or bullish market as evidenced by MSFT's performance in Figure~\ref{fig1:four_stock_comparison}, exhibited increased volatility and a decline in the face of a market downturn. The risk-averse profile, on the other hand, maintained a more conservative stance, often opting to hold positions. This approach resulted in a Cumulative Return trajectory that generally lagged behind the market baseline, reflecting a degree of overcaution that limited trading activity and potential gains, particularly in a bullish market.

\begin{table}[htbp]
\centering
\begin{tabular}{lcccc}
\toprule
\textbf{Metric} & \textbf{B\&H} & \textbf{Self Adaptive} & \textbf{Risk Seeking} & \textbf{Risk Averse} \\
\midrule
\textbf{Cumulative Return (\%)} & -66.9497 & \textbf{54.6958} & -19.4132 & -12.4679 \\
\midrule
\textbf{Sharpe Ratio} & -2.0845 & \textbf{2.4960} & -0.7866 & -1.5783 \\
\midrule
\textbf{Daily Volatility (\%)} & 3.9527 & 2.7419 & 3.2722 & \textbf{1.7744} \\
\midrule
\textbf{Annualized Volatility (\%)} & 3.8050 & 2.5960 & 2.9236 & \textbf{0.9358} \\
\midrule
\textbf{Max-Drawdown (\%)} & 67.3269 & \textbf{12.5734} & 45.0001 & 15.9882 \\
\bottomrule
\end{tabular}
\caption{Comparison of overall trading performance during the testing period with different risk inclinations setting in \textsc{FinMem}'s profiling module.}
\label{tab:diff_char_table}
\end{table}

\subsection{Impact of adjusting the capacity of \textsc{FinMem} working memory (RQ5)} 

In our third study, we explored whether appropriately tuning the memory retrieval bandwidth of \textsc{FinMem} could enhance its trading performance. This bandwidth is tied to the working memory's capacity within its memory module. As depicted in Figure~\ref{fig1:training_test_logic}, \textsc{FinMem} retrieves the top-$K$ memory events from its long-term memory in response to a trading inquiry. The working memory capacity is thus set at $3 \times K$, mirroring the human cognitive system's limit of processing immediate messages upon specific stimuli ($3$ refers to the three processing layers in long-term memory). By varying the $K$ hyperparameter, \textsc{FinMem} can expand this capacity far beyond the human cognitive scope. We aimed to determine whether such flexibility in adjusting memory bandwidth translates to improvements in \textsc{FinMem}'s performance.

\begin{table}[htbp]
\centering
\renewcommand{\arraystretch}{1.35}
\begin{tabular}{lccccc}
\toprule
\textbf{Metric} & \textbf{B\&H} & \textbf{Top 1} & \textbf{Top 3} & \textbf{Top 5} &\textbf{Top 10} \\
\midrule
\textbf{Cumulative Return (\%)} & -66.9497 & 52.0936 & 29.4430 & 54.6958 & \textbf{79.4448} \\
\midrule
\textbf{Sharpe Ratio} & -2.0845 & 1.8642 & 1.1214 & 2.4960 & \textbf{2.7469} \\
\midrule
\textbf{Daily Volatility (\%)} & 3.8050 & 3.3105 & 3.1105 & \textbf{2.5960} & 3.4262 \\
\midrule
\textbf{Annualized Volatility (\%)} & 60.4020 & 52.5529 & 49.3779 & \textbf{41.2100} & 54.3891 \\
\midrule
\textbf{Max Drawdown (\%)} & 67.3269 & 25.2355 & 27.0972 & \textbf{12.5734} & 17.1360 \\
\bottomrule
\end{tabular}
\caption{Comparison of overall trading performance during the testing period with different configurations of working memory capacity.}
\label{tab:topKmem}
\end{table}

As demonstrated in Table~\ref{tab:topKmem}, we adjusted the hyperparameter $K$ to alter the number of memory events retrieved from shallow, intermediate, and deep long-term memory layers in \textsc{FinMem}. We tested $K$ values of 1, 3, 5, and 10, exploring \textsc{FinMem}'s working memory capabilities at levels below, near, and above the human cognitive limit. For all these $K$ settings, we maintained a self-adaptive risk inclination, while other settings were consistent with those described in Section~\ref{param_settings}.

Across all $K$ configurations, \textsc{FinMem} outperformed the Buy \& Hold baseline, indicating the effectiveness of its memory module in processing diverse information and capturing critical events, which subsequently enhanced its trading performance, as evidenced by positive Cumulative Returns and Sharpe Ratios. Notably, higher $K$ values, like 5 and 10, enabled \textsc{FinMem} to achieve the best Cumulative Returns and Sharpe Ratios exceeding 2.0. With $K$ set to 1, \textsc{FinMem} still performed moderately well by capturing the most critical memory events of each layer. 

\begin{figure} 
    \centering
    \includegraphics[width=0.88\linewidth]{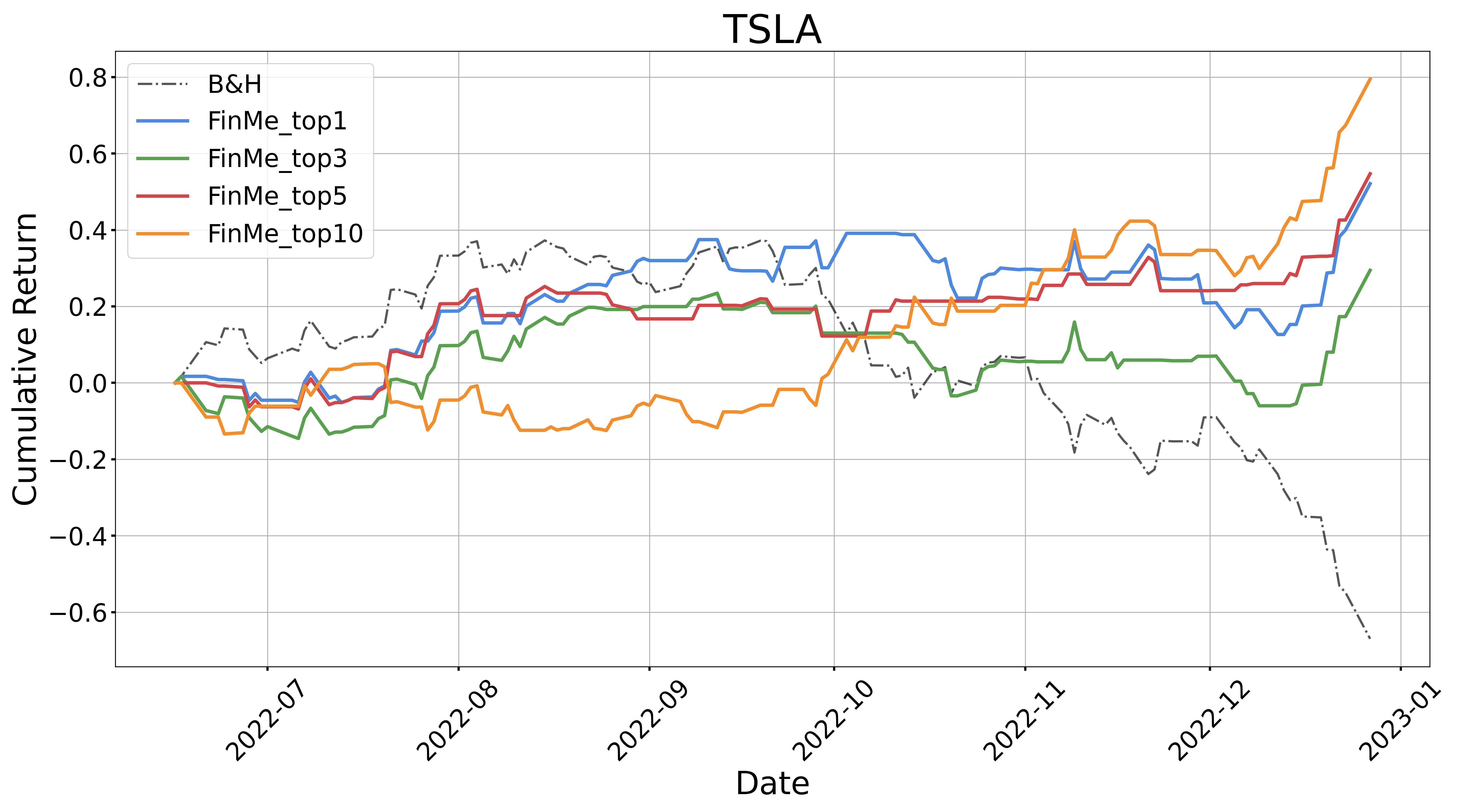}
  \caption{Cumulative Return over time for with different \textsc{FinMem} working memory capacity.}
  \label{fig5:topM} 
\end{figure}

An in-depth analysis in Figure~\ref{fig5:topM}, which shows the Cumulative Return over time for various $K$ settings, reveals that a $K$ value of 5 is optimal for trading TSLA stock, consistently delivering robust performance with the lowest Volatility and Max-Drawdown. Before mid-October 2022, when the stock market was relatively stable and slightly upward, \textsc{FinMem}'s trading actions aligned well with market trends (referring to B\&H) and avoided significant losses. During periods of high volatility and continuous downturns (post-mid-October 2022), it maintained earnings by reducing “Buy” actions and favoring more “Hold” and “Sell” strategies. However, setting $K$ to 10, while effective during market volatility, resulted in significant losses in stable market conditions. The issue may stem from the disproportionately loose capacity constraints on incoming information relative to the volume of incoming data. The broad memory retrieval bandwidth might have mixed trivial messages with critical ones, hampering \textsc{FinMem}'s decision precision.
This challenge becomes especially evident in neutral market conditions, where the influx of information includes a mix of varying market sentiments and trends. 

Addressing \textbf{RQ5}, appropriately tuning the number of memory events (Top-$K$ in the \textsc{FinMem} memory module can significantly enhance its trading performance. The aforementioned study illustrates that \textsc{FinMem} can achieve optimal results by effectively assimilating key signals from a sufficient quantity of filtered memories across each layer. However, the optimal value for $K$ may vary depending on the volume and quality of incoming information.

\section{Conclusion and future work}

In this paper, we introduce \textsc{FinMem}, an innovative automated trading agent framework featuring an adjustable cognitive memory structure and dynamic character design. Our research demonstrates its capacity to enhance stock trading performance substantially using real-world financial datasets. Additionally, the efficacy of each critical component within \textsc{FinMem} is thoroughly demonstrated in our ablation studies, highlighting their roles in optimizing trading outcomes. 

 Its unique features of human-like cognitive memory modules and dynamic character design enable it to tackle the complexities of financial environments and respond aptly to new situations. Compared to other LLM trading agents, \textsc{FinMem}'s memory module equips it with the capability to better comprehend financial data featured by various timeliness and organize them as a self-evolving long-term memory layer. The dynamic character design endows \textsc{FinMem} with critical professional insights, enabling efficient filtering of impactful messages from incoming financial data for trading actions. Additionally, the integration of multiple risk profiles enhances \textsc{FinMem}'s adaptability to a range of market conditions. 

\textsc{FinMem}'s exceptional performance underscores its remarkable ability to transform diverse financial data into well-informed investment strategies. Its proficiency in integrating various financial data types is further accentuated by a notably reduced training duration, which offers advantages for trading with newly established companies. In our approach, we utilized a limited range and quality of financial news and reports, employing general-purpose LLMs as the backbone algorithms. However, \textsc{FinMem} is fully compatible with LLMs specifically fine-tuned for financial applications. We anticipate that its trading efficacy will be elevated further with access to a more comprehensive and higher-quality dataset, coupled with LLMs tailored specifically for financial contexts.

While primarily designed for financial decision-making, the \textsc{FinMem} framework boasts a versatility that extends to domains such as IT consulting and business reporting, where actions are driven by time-sensitive information. Looking ahead, an intriguing direction for development is the creation of a multi-agent trading system, rooted in the \textsc{FinMem} platform, aimed at enhancing investment portfolio optimization. This system, featuring diverse professional backgrounds in its profiling modules, enables concurrent operations and trading across a variety of financial products. It dynamically adjust trading targets based on a sequential assessment of key trading performance indicators. Simultaneously, it can timely reallocate investment proportions for each financial product category, leveraging peer-to-peer communication and systematic performance analysis. Supported by LLMs, this multi-agent trading system is poised to assemble an optimized investment portfolio through its strategically planned operations.

\printbibliography

\end{document}